\documentclass[letterpaper,twocolumn,10pt]{article}
\usepackage{usenix}
\usepackage{amsmath}
\usepackage{contour,ulem}

\contourlength{0.8pt}

\usepackage[numbers]{natbib} 
\usepackage{tikz}

\usepackage{algorithm,algpseudocode,algorithmicx}
\algnewcommand\algorithmicforeach{\textbf{for each}}
\algdef{S}[FOR]{ForEach}[1]{\algorithmicforeach\ #1\ \algorithmicdo}
\algdef{SE}[DOWHILE]{Do}{doWhile}{\algorithmicdo}[1]{\algorithmicwhile\ #1}
\usepackage{fontawesome}
\usepackage{colortbl} 
\usepackage{booktabs} 
\usepackage{multirow,multicol} 
\usepackage{makecell} 
\usepackage[switch]{lineno} 
\usepackage{threeparttable} 
\usepackage{xspace} 
\usepackage{fancyhdr} 
\usepackage{xurl} 
\DeclareUrlCommand\myurl{\urlstyle{tt}} 
\newcommand*{\doi}[1]{\href{https://doi.org/#1}{\myurl{DOI: #1}}}

\newcommand{\para}[1]{\vspace{0.75mm}\noindent\textbf{#1.}}

\usepackage{adjustbox} 
\newcolumntype{R}[2]{>{\adjustbox{angle=#1,lap=\width-(#2)}\bgroup}l<{\egroup}}

\newcommand{\quotetext}[1]{{\textsuperscript{\tiny\faQuoteLeft}} {\small\textit{#1}} {\textsubscript{\tiny\faQuoteRight}}}

\newcommand{\hackforums}{{\small\scshape Hack Forums}\xspace}

\newcommand{\darkmarket}{{\small\scshape DarkMarket}\xspace}

\newcommand{\ccc}{Cambridge Cybercrime Centre\xspace}

\newcommand{\netscout}{{\scshape Netscout}\xspace}
\newcommand{\hopscotch}{{\scshape Hopscotch}\xspace}
\newcommand{\amppot}{{\scshape AmpPot}\xspace}

\newcommand{\anh}[1]{\{{\textcolor{blue}{anh:~#1}}\}}

\newcommand{\MSnGoogleBotPrefixes}{{1\,163}}
\newcommand{\MSnBingBotPrefixes}{{26}}
\newcommand{\MSnTotalBotPrefixes}{{1\,189}}
\newcommand{\BTnTotalNCATrafficVisitsDuringTwoMonthsWeb}{{7\,001}}
\newcommand{\BTnTotalNCATrafficVisitsDuringTwoMonthsPeakWeb}{{1\,234}}

\newcommand{\BTnTotalSimilarwebDomains}{{94}}
\newcommand{\BTsimilarwebCorrelationCoefficient}{{0.81}}
\newcommand{\BTsimilarwebCorrelationCoefficientPValue}{{.001}}
\newcommand{\BTsimilarwebCorrelationCoefficientConfidenceIntervalLow}{{0.76}}
\newcommand{\BTsimilarwebCorrelationCoefficientConfidenceIntervalHigh}{{0.85}}
\newcommand{\BTnGroundTruthSessionsWithSingleRequest}{{832k}}
\newcommand{\BTnGroundTruthSessionsWithSingleRequestProps}{{47.09}}
\newcommand{\BTnGroundTruthSessionsWithMultipleRequests}{{935k}}
\newcommand{\BTnGroundTruthSessionsWithMultipleRequestsProps}{{52.91}}
\newcommand{\BTnGroundTruthTopDomainsProps}{{40.57}}
\newcommand{\BTnGroundTruthTopDomainsipstresserxcomProps}{{12.98}}
\newcommand{\BTnGroundTruthTopDomainsinstantstresserxcomProps}{{9.45}}
\newcommand{\BTnGroundTruthTopDomainsstresserxappProps}{{7.29}}
\newcommand{\BTnGroundTruthTopDomainsbootyouxnetProps}{{6.36}}
\newcommand{\BTnGroundTruthTopDomainsfreestresserxsoProps}{{4.49}}
\newcommand{\BTnGroundTruthTopCountriesProps}{{70.68}}
\newcommand{\BTnGroundTruthTopCountriesUSProps}{{37.43}}
\newcommand{\BTnGroundTruthTopCountriesCNProps}{{5.51}}
\newcommand{\BTnGroundTruthTopCountriesDEProps}{{5.04}}
\newcommand{\BTnGroundTruthTopCountriesGBProps}{{4.55}}
\newcommand{\BTnGroundTruthTopCountriesRUProps}{{4.09}}

\newcommand{\BTnGroundTruthTopBrowserChromeProps}{{34.61}}
\newcommand{\BTnGroundTruthTopBrowserFirefoxProps}{{14.90}}

\newcommand{\BTnGroundTruthTopOSWindowsProps}{{48.59}}
\newcommand{\BTnGroundTruthTopOSMacOSXProps}{{14.19}}

\newcommand{\BTnGroundTruthTopDeviceTypePCProps}{{77.63}}
\newcommand{\BTnGroundTruthTopDeviceTypeMobileProps}{{20.42}}
\newcommand{\BTnGroundTruthTopDeviceTypeTabletProps}{{1.95}}
\newcommand{\BTnTotalBotTypes}{{237}}
\newcommand{\BTnTotalRawEvent}{{20.7M}}
\newcommand{\BTnTotalVisitSessions}{{2.7M}}
\newcommand{\BTnrealvisitVisits}{{1.8M}}
\newcommand{\BTnrealvisitVisitsProps}{{66.46}}
\newcommand{\BTnbotVisits}{{816.4k}}
\newcommand{\BTnbotVisitsProps}{{30.71}}
\newcommand{\BTnapicallVisits}{{70.5k}}
\newcommand{\BTnapicallVisitsProps}{{2.65}}
\newcommand{\BTnprefetchVisits}{{4.9k}}
\newcommand{\BTnprefetchVisitsProps}{{0.18}}

\newcommand{\BTnFirstWaveSeizedBooters}{{48}}
\newcommand{\BTnFirstWaveSeizedDomains}{{49}}
\newcommand{\BTnFirstWaveSeizedDomainsThatHasSimilarwebTraffic}{{49}}

\newcommand{\BTnFirstWaveResurrectedBooters}{{25}}
\newcommand{\BTnFirstWaveResurrectedDomainsThatHasSimilarwebTraffic}{{31}}
\newcommand{\BTnFirstWaveGaveUpBooters}{{23}}

\newcommand{\BTnSecondWaveSeizedBooters}{{11}}
\newcommand{\BTnSecondWaveSeizedDomains}{{13}}
\newcommand{\BTnSecondWaveSeizedDomainsThatHasSimilarwebTraffic}{{13}}

\newcommand{\BTnSecondWaveResurrectedBooters}{{11}}
\newcommand{\BTnSecondWaveResurrectedDomainsThatHasSimilarwebTraffic}{{11}}

\newcommand{\BTnSecondWaveResurrectedTwiceBooters}{{9}}

\newcommand{\BTnMeanResurrectionTwiceCombinedBothWave}{{39}}
\newcommand{\BTnMedianResurrectionTwiceCombinedBothWave}{{27}}

\newcommand{\BTnMeanReinstallationTwiceCombinedBothWave}{{8}}
\newcommand{\BTnMedianReinstallationTwiceCombinedBothWave}{{1}}
\newcommand{\BTnCollectedTelegramChannels}{{52}}
\newcommand{\BTnTelegramMessages}{{34\,438}}
\newcommand{\BTnTelegramReplies}{{5\,246}}

\newcommand{\BTnTelegramEmojis}{{6\,290}}

\newcommand{\BTnMeanFirstWaveResurrection}{{66}}
\newcommand{\BTnMedianFirstWaveResurrection}{{19}}
\newcommand{\BTnMeanFirstWaveReinstallation}{{59}}
\newcommand{\BTnMedianFirstWaveReinstallation}{{8}}
\newcommand{\BTnMeanSecondWaveResurrection}{{50}}
\newcommand{\BTnMedianSecondWaveResurrection}{{42}}
\newcommand{\BTnMeanSecondWaveReinstallation}{{11}}
\newcommand{\BTnMedianSecondWaveReinstallation}{{2}}
\newcommand{\BTnFirstWaveUSRegistrar}{{43}}
\newcommand{\BTnFirstWaveUSRegistrarProps}{{88}}
\newcommand{\BTnFirstWaveNameCheapRegistrar}{{40}}
\newcommand{\BTnFirstWaveNameCheapRegistrarProps}{{82}}
\newcommand{\BTnSecondWaveUSRegistrar}{{9}}
\newcommand{\BTnSecondWaveUSRegistrarProps}{{69}}
\newcommand{\BTnSecondWaveNameCheapRegistrar}{{7}}
\newcommand{\BTnSecondWaveNameCheapRegistrarProps}{{54}}
\newcommand{\BTnHighlyRelevantThreads}{{74}}
\newcommand{\BTnHighlyRelevantPosts}{{927}}
\newcommand{\BTnLowerRelevantThreads}{{365}}
\newcommand{\BTnLowerRelevantPosts}{{777}}
\newcommand{\BTnAllRelevantPosts}{{1\,704}}
\newcommand{\BTnAllRelevantUsers}{{714}}
\newcommand{\BTnHopscotchRecords}{{4.6M}}

\newcommand{\BTnAmpPotRecords}{{9.8M}}

\newcommand{\BTnUDPandTCPNetscoutRecords}{{32.9M}}
\newcommand{\BTnSelfReportedBooters}{{207}}
\newcommand{\BTGroundTruthSessionsUsingProxy}{{61k}}
\newcommand{\BTGroundTruthSessionsUsingProxyProps}{{2.28}}
\newcommand{\BTGroundTruthRequestsUsingProxy}{{616k}}
\newcommand{\BTGroundTruthRequestsUsingProxyProps}{{2.97}}

\newcommand{\BTGroundTruthRequestsUsageTypeDCHProps}{{97.34}}

\newcommand{\BTGroundTruthRequestsUsageTypeISPProps}{{1.78}}

\newcommand{\BTNavigationCountsFromipstressercom}{{29\,090}}

\newcommand{\BTNavigationCountsToinstantstressercom}{{37\,690}}

\newcommand{\BTNavigationCountsFromipstressercomToinstantstressercom}{{10\,906}}
\newcommand{\BTNavigationCountsFromipstressercomTofreestresserso}{{8\,579}}
\newcommand{\BTNavigationCountsFrominstantstressercomTostresserapp}{{8\,760}}
\newcommand{\BTNavigationCountsFrombootyounetToinstantstressercom}{{11\,704}}
\newcommand{\BTNavigationCountsFromfreestressersoToinstantstressercom}{{9\,272}}
\newcommand{\BTNavigationCountsFrominstantstressercomTostresseraicom}{{6\,592}}
\newcommand{\BTNavigationCountsFromNCADomains}{{6\,849}}
\newcommand{\BTNavigationCountsToNCADomains}{{7\,289}}
\newcommand{\BTNavigationCountsWithinAllNCADomains}{{729}}
\newcommand{\BTNIPAccessingDomainsTotal}{{653\,107}}

\newcommand{\BTNIPAccessingOnlyOneDomainProps}{{75.04}}

\newcommand{\BTNIPAccessingMultipleDomainProps}{{24.96}}

\newcommand{\BTNIPAccessingMoreThanFiveDomainProps}{{3.16}}

\newcommand{\BTNIPAccessingMoreThanTenDomainProps}{{0.92}}

\newcommand{\BTNApiCallSessionsTopTenProps}{{97.01}}
\newcommand{\BTNApiCallingUsersTopTen}{{9\,427}}
\newcommand{\BTNApiCallingUsersTopTenProps}{{92.87}}
\begin{document}

\setlength{\headheight}{15pt} 
\setlength{\headsep}{24pt}  
\addtolength{\topmargin}{-14.5mm}

\title{\Large \bf Assessing the Aftermath: the Effects of a Global Takedown\\against DDoS-for-hire Services}
\author{
    {\rm Anh V. Vu}\\University of Cambridge\\anh.vu@cl.cam.ac.uk\vspace{3mm}\\
    {\rm John Kristoff}\\University of Illinois Chicago\\jkrist3@uic.edu\vspace{4mm}
    \and
    {\rm Ben Collier}\\University of Edinburgh\\ben.collier@ed.ac.uk\vspace{3mm}\\
    {\rm Richard Clayton}\\University of Cambridge\\richard.clayton@cl.cam.ac.uk\vspace{4mm}
    \and
    {\rm Daniel R. Thomas}\\University of Strathclyde\\d.thomas@strath.ac.uk\vspace{3mm}\\
    {\rm Alice Hutchings}\\University of Cambridge\\alice.hutchings@cl.cam.ac.uk\vspace{4mm}
}
\maketitle

\begin{abstract}
Law enforcement and private-sector partners have in recent years conducted various interventions to disrupt the DDoS-for-hire market. Drawing on multiple quantitative datasets, including web traffic and ground-truth visits to seized websites, millions of DDoS attack records from academic, industry, and self-reported statistics, along with chats on underground forums and Telegram channels, we assess the effects of an ongoing global intervention against DDoS-for-hire services since December 2022. This is the most extensive booter takedown to date conducted, combining targeting infrastructure with digital influence tactics in a concerted effort by law enforcement across several countries with two waves of website takedowns and the use of deceptive domains. We found over half of the seized sites in the first wave returned within a median of one day, while all booters seized in the second wave returned within a median of two days. Re-emerged booter domains, despite closely resembling old ones, struggled to attract visitors (80--90\% traffic reduction). While the first wave cut the global DDoS attack volume by 20--40\% with a statistically significant effect specifically on UDP-based DDoS attacks (commonly attributed to booters), the impact of the second wave appeared minimal. Underground discussions indicated a cumulative impact, leading to changes in user perceptions of safety and causing some operators to leave the market. Despite the extensive intervention efforts, all DDoS datasets consistently suggest that the illicit market is fairly resilient, with an overall short-lived effect on the global DDoS attack volume lasting for at most only around six weeks. 
\end{abstract} 

\section{Introduction} \label{sec:introduction}
The economic and social dynamics of the cybercrime ecosystem have fundamentally changed over the past two decades. There have been significant shifts in both the scale and nature of harmful and illicit activity resulting from the increasing industrialisation of economic cybercrime. Individuals using simple methods remain an issue, but they now have the option of using fully-fledged commercial service offerings. Off-the-shelf toolkits and attack-as-a-service infrastructure are routinely advertised, rented, and sold through a variety of open and private online marketplaces~\cite{akyazi2021measuring,hyslip2020cybercrime,collier2021cybercrime}. Each stage of this evolution has drastically reduced the skill and cost barriers for users to participate in what were previously considered more `technical' forms of online harm~\cite{karami2013understanding}.

The as-a-service model facilitates largely underreported high-volume but low-value online crimes, which may avoid public scrutiny and law enforcement attention. For example, starting at a few dollars per month, online markets for DDoS attacks (so-called \textit{booters} or \textit{stressers}) allow unskilled actors to flood online services with unwanted traffic, knocking systems without robust security offline~\cite{hutchings2016exploring}. Booter operators may justify their services as testing server resilience, but their primary uses and methods are illegal~\cite{douglas2017booters}.

The entrepreneurial, service-based nature of booters makes them vulnerable to targeted interventions. Although individual arrests may be generally ineffective in industrialised market crime economies as they aid competitors, the economic forces driving these markets tend to lead to the centralisation of the infrastructure and supportive work on which they depend (through well-understood efficiencies and network externalities). Booter infrastructure consolidation therefore presents an apposite target for disruption activities. In a major takedown event in December 2018, 15 of the largest booters were seized in an international law enforcement effort~\cite{xmasevent2018}. However, the effect was relatively short-lived, with DDoS attack volumes recovering a few weeks later~\cite{collier2019booting, kopp2019ddos}.

Building on practitioner learning and academic analysis of previous efforts, a further major intervention was conducted from late 2022, combining targeting infrastructure with influence operation components: 49 booter domains were seized in December 2022 (the first wave)~\cite{booterseizure2022}, 13 domains were seized in May 2023 (the second wave)~\cite{seized13moredomains}, and several deceptive sites were set up by law enforcement. This campaign has been much greater in scale and more persistent than the one four years earlier, leading us to ask: how effective might it be?

There are several material aspects of online crime that make it more measurable than comparable forms of `offline' crime -- giving researchers a far more detailed global picture of prevalence and the takedown effects than is possible for most other crime types. However, as different parts of the Internet infrastructure generate rather different (and often partial) views of digital phenomena, producing this picture is complex, requiring analysis of multiple datasets. We combine various measurement sources of DDoS attack data with ground-truth traffic gathered through law enforcement-deployed `splash pages' on seized domains to draw unique insights into how booter users visit and move between services after seizures.

We undertake the first measurement of this takedown effort. We sketch out the DDoS-for-hire market and previous interventions in~\S\ref{sec:ddos-for-hire-market}, then describe our methods in~\S\ref{sec:methods-and-datasets}. We outline the analysed takedowns, resurrections, ground-truth booter characteristics, and how users access and move across booters in~\S\ref{sec:takedown-resurrection}. We subsequently explore the longitudinal effects on booter traffic, DDoS attack volumes, and community perceptions in~\S\ref{sec:takedown-effects}, then discuss implications in \S\ref{sec:discussion}. This work was approved by our department's research ethics committee, see \autoref{appendix:ethical-issues}. Our datasets and scripts can be made available to researchers through various providers, see \autoref{appendix:data-licensing}.

\pagestyle{fancy}
\fancyhead[L]{\ifthenelse{\isodd{\value{page}}}{}{\small{In Proceedings of the USENIX Security Symposium 2025}\vspace{0.05mm}}}
\fancyhead[R]{\ifthenelse{\isodd{\value{page}}}{\small{In Proceedings of the USENIX Security Symposium 2025}\vspace{0.05mm}}{}}
\fancyfoot[L]{\ifthenelse{\isodd{\value{page}}}{\small{Anh V. Vu, Ben Collier, Daniel R. Thomas, John Kristoff, Richard Clayton, and Alice Hutchings}}{\small{\thepage}}}
\fancyfoot[R]{\ifthenelse{\isodd{\value{page}}}{\small{\thepage}}{\small{Anh V. Vu, Ben Collier, Daniel R. Thomas, John Kristoff, Richard Clayton, and Alice Hutchings}}}
\fancyfoot[C]{}

\section{The DDoS-for-hire Market} \label{sec:ddos-for-hire-market}
DDoS-for-hire services are subscription-based websites allowing users to carry out DDoS attacks with little technical skills, even using commands in chat channels. Subscriptions start at a few dollars, or around \$20 for unlimited monthly use~\cite{hutchings2016exploring, karami2013understanding}. Booter-generated attacks are often not powerful enough to shut down major sites, but they can effectively congest home connections or unprotected servers. The attack duration is typically short (5--10 minutes), but multiple sessions can be combined. There are links between the `producer' and `consumer' sides of the market; many booter operators were customers prior to managing their services~\cite{hutchings2016exploring,gatrelarrest}. While they claim legitimacy as tools for stress-testing networks and infrastructure~\cite{krebthinkagain}, booters are predominantly used illegally, e.g., teenagers disrupting online exams at schools~\cite{santanna2015booters} and gaining a competitive edge by booting opponents off online games~\cite{karami2013understanding}. Many users incorrectly believe that using booters will not violate the law~\cite{hutchings2016exploring}, and is not an enforceable offence.

A common vector attributed to booters involves reflective attacks exploiting UDP protocols (e.g., DNS, NTP, LDAP)~\cite{krupp2017linking}, where small IP-spoofed packets sent to misconfigured devices are amplified into larger payload responses to victims. UDP attacks are straightforward, not relying on an open port but on exhausting bandwidth capacity~\cite{santanna2015booters}. It is challenging to block and attribute sources, but modern sites can be guarded by third-party security layers such as Cloudflare and DDoS-Guard that filter and drop malicious packets, making UDP-based attacks less effective.\footnote{~Many booters also deploy this protection to conceal their hosting from the public and avoid being DDoSed by competitors~\cite{santanna2015booters}. But this allows them to be located if those protectors comply with law enforcement requests for information, for example, the arrest of a booter operator in 2018 began with an investigation into his Cloudflare service linked to his Google account~\cite{gatrelcase}.} Direct-path attacks, such as SYN and ACK flooding, are also used. Many booters also offer attacks at the application layer with a three-way handshake completed before executing application-layer commands e.g., fetching large files. A multitude of compromised machines (e.g., a botnet) can be used to flood requests, making them appear legitimate and challenging to filter out.

Empirical studies analysing publicly-leaked databases and test purchases from booters show that booters are responsible for hundreds of thousands of DDoS attacks annually~\cite{karami2016stress, karami2013understanding}, some of which cause significant harm~\cite{santanna2015booters}. While revenues for most booters are trivial, some large ones made thousands of dollars a month~\cite{karami2013understanding} and hundreds of thousands in total~\cite{titaniumstresserarrest, vdosleak}. The revenue and profit appear to well sustain their business and maintain the infrastructure, estimated by the advertised price and number of users~\cite{smirnova2024exploring}. Payment methods have, due to interventions, shifted from credit cards~\cite{vdospaypalintervention} to cryptocurrency, thus, limiting access to convenient payments can be a disruptive mechanism~\cite{karami2016stress,brunt2017booted}. Running booters is illegal in most jurisdictions, with operators arrested and jailed in the US, UK, and Netherlands~\cite{collier2019booting}. Using booters is also against the law and may result in criminal liability, but law enforcement tend to pursue operators instead of users -- whom they believe to largely be minors, and hence prevention is more appropriate than pursuit~\cite{hutchings2016exploring}.

\subsection{The Impact of Prior Interventions} \label{subsec:prior-interventions}
There have been direct and indirect attempts by the industry and community to disrupt booter services. In 2017, PayPal shut down booter-associated accounts, causing significant drops in their revenue and prompting them to switch to crypto payments~\cite{brunt2017booted}. In October 2016, the primary booter-related section on \hackforums, `Server Stress Testing', was dismantled~\cite{booterbannedhackforum}. This (at the time) largest hacking forum also banned booter adverts, leading to the emergence of a more dispersed customer ecosystem of Discord and Telegram channels run to promote individual services, with no central community site. Booters also promoted themselves with adverts, which was against the policies of most search engines~\cite{claytongooglebooter}.

Law enforcement has also influenced the market in various ways. Three arrests were announced in a global takedown in December 2018; 15 domains were seized, immediately terminating eight booters~\cite{xmasevent2018}. Law enforcement generally time interventions at Christmas, due to increased gaming activity (and DDoS attacks) over the holidays. This campaign significantly reduced both UDP amplification and self-report DDoS attack counts for a few weeks~\cite{collier2019booting}, as also reported in a separate study using booters for self-attack and analysing traffic from ISPs and Internet Exchange Points (IXPs)~\cite{kopp2019ddos}.

Prior work evaluated a UK National Crime Agency (NCA) influence campaign using Google ads from December 2017 to June 2018. The ads were delivered within the UK to target and warn those searching for DDoS-related terms about the illegality and potential criminal liability of DDoS attacks. This approach appeared effective: only a few thousand pounds spent on ads flattened the demand for DDoS attacks targeting the UK for six months, while attacks on other countries continued to rise. However, the apparent effect vanished shortly after the ads discontinued~\cite{collier2019booting}. These targeted ads were subsequently deployed in other European countries, with mixed short-term effects in reducing the DDoS attack volume~\cite{moneva2023effect}. 

There have been other concerted law enforcement efforts to disrupt booter services over almost a decade. Much of the historical interventions have been described and evaluated in prior work~\cite{collier2019booting,bada2023evaluation}. These interventions have included takedowns, arrests, and warnings delivered to registered users.

Prior interventions caused considerable declines in booter activity, but the effects were often short-lived, with displacement to other booters occurring. A qualitative analysis of cybercrime forums suggests minor deterrent effects of law enforcement actions on users, who claimed to move to other sites, desisted, or changed their practices as a result~\cite{decary2023like}.

\subsection{The Ongoing Interventions} \label{subsec:the-recent-takedown}
The global takedown we analyse, involving the FBI, NCA, and Dutch Police, commenced four years after the last major event (Operation PowerOFF).\footnote{~Operation PowerOFF: https://operation-poweroff.com/} This was carried out in collaboration with academics (some of the authors) and private industry; working booters were monitored weekly, tested and ranked, while new domains and self-reported statistics were recorded. Law enforcement also conducted investigations, using test purchases to determine if services were credibly involved in illegal activities. The intention was to support the takedown of the largest booter, along with as many other capable and visible booters as possible. Relevant hosting providers and domain registrars within the US, such as Cloudflare and Namecheap, then assisted in suspending and redirecting seized domains to a police-deployed page to collect access information. This is the most extensive booter takedown conducted so far.

The 1\textsuperscript{st} wave began on 14 December 2022, with 49 domains of the 48 largest booters seized by the FBI~\cite{booterseizure2022}, taking offline half of the booters operational at that time.\footnote{~Booters often have multiple domains and some depend on the infrastructure of larger services, operating by using API calls provided by others.} Six individuals in the US were arrested and charged for their alleged involvement in operating websites offering Denial of Service attacks, while an 18-year-old operator in the UK was also arrested by the NCA. Shortly after the 1\textsuperscript{st} wave, a major booter dashboard ceased operations. In Germany, a hosting provider offering services to cybercrime-related infrastructure, including DDoS-for-hire, was seized in late March 2023~\cite{germanseizedbooters}. The 2\textsuperscript{nd} wave occurred on 5 May 2023 (four months after the 1\textsuperscript{st} wave), with 13 more domains seized~\cite{seized13moredomains}. This was a notably faster pace of action compared to the previous four-year interval. The German police took down a booter in April 2024, arrested its operators, and seized its databases and servers~\cite{stressertechseizure}. In July 2024, a Texas man was sentenced to 9 months in prison~\cite{texasmansentenced}. In November 2024, a popular DDoS review platform was seized with two arrests in Germany; its Telegram channel was also wiped~\cite{dstatccseized}. Later in mid-December 2024, law enforcement seized 27 booters, arrested three operators, and identified 300 users for further planned actions~\cite{bootertakedown2024}.

Police-controlled services have been used to covertly collect intelligence of other online illegal markets, such as when an influx of users migrated to Hansa market after AlphaBay was seized~\cite{hansainflux}, without knowing that Hansa's server in the Netherlands was monitored by the Dutch Police.\footnote{~The server later moved to Lithuania and was discovered by a payment made from a Bitcoin address recorded in previous logs from the old server.} A decade earlier, the FBI had infiltrated \darkmarket, running the servers from their offices in Pittsburgh~\cite{glenny2011darkmarket}. In October 2024, the NexFundAI token was created by the FBI to investigate crypto market manipulation~\cite{fbinexfundai}. These approaches raise some legal and ethical questions around the possibility of entrapment or iatrogenic harm. This time, the NCA employed similar approaches towards a different end -- setting up honeypot services to which customers were intended to migrate after the takedowns. Rather than collect intelligence, the intention was to weaken trust in the market and increase perceptions of risk for users. This influence campaign was deployed in anticipation of the 1\textsuperscript{st} wave, by setting up deceptive sites appearing to be legitimate booters. With many booters down, these deceptive ones attracted users to register accounts, then displayed warnings to inform them of the operations. They aimed to identify and educate potential customers about the legal consequences of purchasing and carrying out DDoS attacks. A limited number of in-person visits were undertaken to registered users to deliver warnings. The existence of these domains was publicly revealed on 24 March 2023~\cite{ncarevealsfakebooters}.

Informed by prior research, the NCA and Dutch Police also purchased search engine adverts targeting users searching for booters, informing them that these activities are illegal.\footnote{~Advertisements promoting deceptive sites may not be welcomed by search engines, even when run by law enforcement. Sponsoring them to appear at the top of search results therefore may not be always straightforward.} This tactic was found to be associated with reduced DDoS attacks targeting the UK for six months in 2018~\cite{collier2019booting}. To censor booters from search results, the NCA has submitted at least 25 government removal requests to Google between April and August 2024~\cite{ncalumendatabase}, which appears to have been effective as booters have now almost disappeared from search engines. The NCA actively monitored and posted on \hackforums, replying to users asking for cybercrime tools and services, including DDoS attacks. The Dutch Police visited several chat channels, including Telegram, to notify booter users about the operation, seized databases, and follow-up actions.

\begin{table}[t]
\centering
\caption{The quantitative data sources used and their origins.}
\setlength{\tabcolsep}{0.32em}\vspace{2mm}
\small
\begin{tabular}{lrrr}
    \toprule
    Datasets & Statistics & Origins \\
    \midrule
    Ground-truth traffic~\textsuperscript{$\diamond$} & \BTnTotalRawEvent~raw events & Our collection\\
    Similarweb analytics~\textsuperscript{$\star$} & \BTnTotalSimilarwebDomains~booter domains & Our collection\\
    \midrule
    \hopscotch~\textsuperscript{$\dagger$} & \BTnHopscotchRecords~records & Thomas et al.~\cite{thomas20171000}\\
    \amppot~\textsuperscript{$\dagger$} & \BTnAmpPotRecords~records & Kr{\"a}mer et al.~\cite{kramer2015amppot}\\
    \netscout~\textsuperscript{$\dagger$} & \BTnUDPandTCPNetscoutRecords~records & \netscout~\cite{netscoutstats}\\
    Self-reported statistics~\textsuperscript{$\dagger$} & \BTnSelfReportedBooters~booters & Our collection\\
    \midrule
    Underground forums~\textsuperscript{$\star$} & \BTnAllRelevantPosts~posts & Pastrana et al.~\cite{pastrana2018crimebb}\\
    Chat channels~\textsuperscript{$\star$} & \BTnTelegramMessages~messages & Our collection\\
    \bottomrule
\end{tabular}
\\{\vspace{1mm}\raggedright 
\footnotesize{Data timespans covering both waves: \textsuperscript{$\diamond$} [14 December 2022 -- 31 July 2023]; \textsuperscript{$\star$} [1 October 2022 -- 30 September 2023]; \textsuperscript{$\dagger$} [1 July 2021 -- 30 June 2023]} \par}
\label{tab:data-sources}
\end{table}
\section{Methods and Datasets} \label{sec:methods-and-datasets}
As part of our collaboration with law enforcement, we regularly monitored all active booters, tested their functionality, collected their self-reported statistics every Monday, and ranked the largest ones based on average daily attack counts prior to takedowns. We also analysed their Telegram channels to ascertain the presence and timing of any resurrected domains (see more below). Additional quantitative datasets are shared with us by other academics and industry. We measured these datasets longitudinally and statistically, with online discussions further being qualitatively analysed to understand community perspectives. A dataset summary is provided in~\autoref{tab:data-sources}. All timestamps in our analyses are normalised to UTC.

\subsection{Traffic to Booter Domains}
Seized domains were redirected to a landing page hosted by us with Cloudflare serverless, displaying messages about the takedown and legal status of booting. All traffic to this page is logged, giving us a complete view on visits to seized domains and additionally, once they too were taken down, the deceptive NCA-deployed domains. As the seizure affected the largest booters in the market, these ground-truth insights account for a large proportion of the users and attacks in the booter ecosystem. However, the term `ground truth' refers to traffic to booter domains after seizure, not during their normal operations. To estimate earlier levels of traffic to the booter domains, traffic to those not seized, and traffic to the domains that emerged after the takedowns, we additionally collected analytics from Similarweb -- a platform providing intelligence into web traffic and performance. As our collections cover the largest booters, these two datasets provide a reasonably good view of the ecosystem. This allows us to garner novel insights into two as-yet previously hidden aspects, namely, how major booters provide API services to smaller ones, and how users visit and move across domains on seizure.

\para{Ground-truth Traffic} To clean the data, API calls, link prefetching, and traffic originating from our equipment were removed. Search engine crawlers were excluded, including Google, Bing, Yahoo, Baidu, Yandex, Sogou and DuckDuckGo, which occupies around 98\% of all spiders' requests (Google dominates with 80--85\%)~\cite{searchenginemarketshare}. We excluded \MSnTotalBotPrefixes~official bot prefixes: \MSnBingBotPrefixes~of Bing and \MSnGoogleBotPrefixes~of Google. Otherwise, we identified web spiders by user agents, which is an imperfect heuristic, but we believe at-scale analysis is indicative of overall behaviour. Web analytics and monitoring services (e.g., SemrushBot, AhrefsBot, SiteChecker, UptimeRobot), Internet archives (e.g., archive.org), and social media spiders (e.g., DiscordBot, TwitterBot and FacebookBot) were also excluded. Metadata (e.g., browser, OS, device type and family) were extracted from user agents, while the originating country of visits was determined by Cloudflare using MaxMind's geolocation database updated twice a week~\cite{cloudflareipgeo}. MaxMind claims to provide over 99.8\% country-level accuracy~\cite{maxmindaccuracy}.

We use IP addresses to distinguish visitors for practical reasons, which has limitations as a single real user may appear under multiple IP addresses, and a public IP address might be assigned to different devices (e.g., behind a NAT router). One can use a specific timeframe (e.g., within 24 hours) to consider IP reuse as the same visitor, but this creates tension between false positives and false negatives and cannot completely eliminate either. There is no easy solution; we believe the overall trend remains indicative, given the data scale of millions of records. A session is then determined by grouping subsequent requests from an IP to a domain until 30 minutes of inactivity (timestamped as the first request), similar to methods used by popular analytics such as Google, Similarweb, and Semrush~\cite{similarwebsessioncount,semrushsessioncount}. In total, this dataset spans from 14 December 2022 (the 1\textsuperscript{st} wave) to 31 July 2023 (three months after the 2\textsuperscript{nd} wave, ending in early August 2023 when domain registrations lapsed), with \BTnTotalRawEvent~raw events (\BTnTotalVisitSessions~sessions) recorded. We excluded \BTnTotalBotTypes~types of bots with \BTnbotVisits~sessions (\BTnbotVisitsProps\%), and analysed \BTnrealvisitVisits~ordinary sessions (\BTnrealvisitVisitsProps\%), \BTnapicallVisits~API call sessions (\BTnapicallVisitsProps\%) and \BTnprefetchVisits~link prefetch sessions (\BTnprefetchVisitsProps\%).

\para{Web Traffic Analytics} Instead of precise counts, Similarweb aggregates anonymous statistics from various sources including their own analytics tools, data shared by partners such as ISPs, crawled sites, and other tracking services. The numbers provided constitute inferential statistics based on heuristic extrapolation of collected data to full Internet scale. We test the reliability of this dataset through comparison with our ground-truth visits to seized domains from the first wave onward. Delays in deploying splash pages in the first two days of takedown led to unstable counts, thus the first two days are excluded from our comparison. Although Similarweb estimated figures are lower, they reflect a highly reliable pattern with a strong positive \textit{Pearson} correlation with our non-bot ground-truth traffic to seized and NCA domains (\textit{r = \BTsimilarwebCorrelationCoefficient}, 95\% CI [\BTsimilarwebCorrelationCoefficientConfidenceIntervalLow, \BTsimilarwebCorrelationCoefficientConfidenceIntervalHigh], $p<\BTsimilarwebCorrelationCoefficientPValue$). This gives us a reasonably reliable indicator of historical trends of traffic to seized domains, newly resurrected domains, and NCA domains.

We used a free account to collect complete statistics from 1 October 2022 (2 months before the 1\textsuperscript{st} wave) to 30 September 2023 (6 months after the 2\textsuperscript{nd} wave), including counts of visits, unique visitors, visit duration, pages per visit, bounce rate, and page views of \BTnFirstWaveSeizedDomainsThatHasSimilarwebTraffic~first-wave seized domains (of \BTnFirstWaveSeizedBooters~booters), \BTnSecondWaveSeizedDomainsThatHasSimilarwebTraffic~second-wave seized domains (of \BTnSecondWaveSeizedBooters~booters), \BTnFirstWaveResurrectedDomainsThatHasSimilarwebTraffic~first-wave resurrected domains (of \BTnFirstWaveResurrectedBooters~booters), \BTnSecondWaveResurrectedDomainsThatHasSimilarwebTraffic~second-wave resurrected domains (of \BTnSecondWaveResurrectedBooters~booters), and all NCA domains.

\subsection{DDoS Attack Records}
As academic and industry measurements rely on different approaches and hence generate distinctive views~\cite{hiesgen2024age,nawrocki2023sok}, we use four separate datasets for a more complete and reliable analysis. We removed an anomaly in the \hopscotch and \amppot data by excluding SSDP attacks on Brazil in July 2022 due to unaggregated horizontal attacks (aka. prefix or carpet-bombing). All DDoS datasets span two years from 1 July 2021 to 30 June 2023, covering both waves of takedown.

\para{The \hopscotch Dataset} UDP protocols are often exploited for reflective DDoS attacks; many are attributed to booters~\cite{kramer2015amppot,santanna2015booters,krupp2017linking}. We use a DDoS dataset collected through a global honeypot set up by the \ccc in 2014~\cite{thomas20171000}, which imitates UDP protocols susceptible to reflective attacks, recording events when attackers scan for reflectors without forwarding amplified traffic to victims. The nature of reflective attacks means victims are known but attack origins are unavailable. This data provides a partial view of the global DDoS attacks, focusing on booter-generated traffic while direct-path attacks such as SYN- and ACK-flooding are not included. The honeypot captures only Layer 4 attacks, with a flow considered an attack if any sensor receives 5 or more packets targeting the same IP or IP prefix. The attack duration is deemed to from the first to the last packet preceding 15 minutes without further packets received. 

This data was used by academics to assess the impact of the 2018 takedown~\cite{collier2019booting} and analyse attacks carried out by low-level cybercrime actors in armed conflict~\cite{vu2024getting}. The data accuracy relies on the sensors' geolocation and the amount of traffic received. Attackers may send packets solely to test working reflectors and may avoid sending IP-spoofed requests to the sensors, which could result in a small fraction of captured traffic not originating from real booters. However, its global-scale trends appear to be indicative and have good coverage. In total, we analysed \BTnHopscotchRecords~such DDoS attack records.

\para{The \amppot Dataset} We use another UDP amplification DDoS attack dataset collected through the \amppot honeypot~\cite{kramer2015amppot}. Similar to \hopscotch, \amppot emulates various UDP protocols vulnerable to DDoS abuse, such as NTP, DNS, SSDP, LDAP, and CHARGEN~\cite{rossow2014amplification}. It has 21 instances deployed across the US, Australia, Brazil, Ireland, Iceland, Japan, Singapore, Greece, and the Netherlands. \amppot actively monitors ports for incoming UDP packets, capturing hundred of thousands records every month. The majority of observed attacks are short-lived, originating from single source (e.g., booters) with most victims experiencing only one attack~\cite{kramer2015amppot}. We analysed \BTnAmpPotRecords~such DDoS attack records.

\para{The \netscout Dataset} Both \hopscotch and \amppot only cover UDP amplification attacks. To provide a wider reliable view, we use another dataset shared by \netscout, a provider of network and application performance management, including DDoS mitigation services. \netscout covers not only booter-generated but also TCP-based and direct-path attacks such as SYN-, ACK-, and GRE-flooding. Their system receives attack alerts from global customers elected to enable feedback sharing, observing millions of alerts in the first half of 2023 with over 45\% lasting between 5-15 minutes~\cite{netscoutstats}. 

Alert feedback is provided only for events at medium or higher severity levels, and alert-associated IP addresses may be anonymised based on customer settings. While alerts are based on NetFlow, they are a higher-level aggregated summary of flows associated with attacks, including start and end times, (anonymised) sources and destinations, protocols, ports, etc. \netscout's observation is at large aggregates of traffic flow at the borders between networks, whereas academic datasets are derived from the standpoint of DDoS participant nodes and infrastructure. Target IP addresses are characterised through local enrichment (to the customer) and separate central enrichment upon data ingestion by \netscout. We use the latter for analysis, with \BTnUDPandTCPNetscoutRecords~aggregated counts over the period covering both TCP- and UDP-based attacks; however, this lacks a by-country view. While \netscout covers more attack vectors, the 5--10 minute duration (and low impact) offered by booters may be insufficient to trigger medium or high alerts. Thus, some booter-related attacks may be overlooked or complicatedly mixed with other uninterrupted vectors.

\para{Self-reported Statistics} To attract users, about three-quarters of booters self-report their number of offered services, active customers, successful attacks, amount of generated traffic, and attack power on their websites. These figures encompass all attack types and locations, including both reflective and direct-path attacks. Booters might be incentivised to manipulate their records, but suddenly inflated figures are obvious to detect by weekly assessment. Using statistical tests for distribution and heteroskedasticity, prior work suggests that these statistics demonstrate the properties expected for naturally-occurring, rather than synthetically generated data~\cite{collier2019booting}. Over two years, we have visited \BTnSelfReportedBooters~booters every Monday to collect their data, resulting in a one-week interval between data points e.g., the pre-takedown and post-takedown points are from the closest Mondays before and after seizure: 12 and 19 December 2022 in the 1\textsuperscript{st} wave, and 1 and 8 May 2023 in the 2\textsuperscript{nd} wave.

\subsection{Online Chats and Discussions}
To understand the changing perspectives of the community and booter users in the aftermath of the takedown, we collected discussions and chats on various online platforms.

\para{Underground Forums} Forums are a crucial part of the cybercrime ecosystem in allowing subcultural reproduction, sharing practices and tools, and community cultivation. \hackforums has long been the most popular hub for discussions on hacking techniques within the low-level cybercrime community (i.e. not serious organised or state-sponsored crime). While the majority of forum users are not provably involved in crime of any kind, some have faced cybercrime-related prosecutions~\cite{pastrana2018characterizing}. The forum offers a marketplace to exchange cybercrime tools and services~\cite{vu2020turning,marjanov2024breaking}, but has shifted focus from technical topics to get-rich-quick schemes, ponzi scams, currency trading, stalking-assistance technology, and frauds, though some technical aspects still remain~\cite{hughes2023digital}. While (as with most public forums) no longer being a hub for technically-driven cybercrime, it still serves as a social space with discussions being a useful signal for the perceptions of the wider ecosystem. Cybercrime-as-a-service models predominate here, but the use of cyberattacks is discouraged. A booter-specific section was closed and the sale of booters was banned~\cite{collier2019booting}, but general discussion and perceptions on this matter remain.

We extracted all booter-related posts on \hackforums from 1 October 2022 (2 months before the 1\textsuperscript{st} wave) to 30 September 2023 (6 months after the 2\textsuperscript{nd} wave) in the CrimeBB dataset~\cite{pastrana2018crimebb} using a set of case-insensitive keywords \textit{`booter'}, \textit{`stresser'}, \textit{`ddoser'}, \textit{`ddos'}, \textit{`ddos-as-a-service'}, and \textit{`ddos-for-hire'} (which was suggested to be comprehensive~\cite{santanna2015booters}), along with two keywords particularly related to the takedowns: \textit{`fbi'} and \textit{`nca'}. We extracted \BTnHighlyRelevantPosts~posts from \BTnHighlyRelevantThreads~highly relevant threads -- those with titles directly containing the keywords. Among the other \BTnLowerRelevantThreads~less-relevant threads, we extracted \BTnLowerRelevantPosts~posts having the keywords. In total, our snapshot includes \BTnAllRelevantPosts~relevant posts made by \BTnAllRelevantUsers~active forum users.

\begin{figure}[t]
    \centering
    \includegraphics[width=0.457\textwidth]{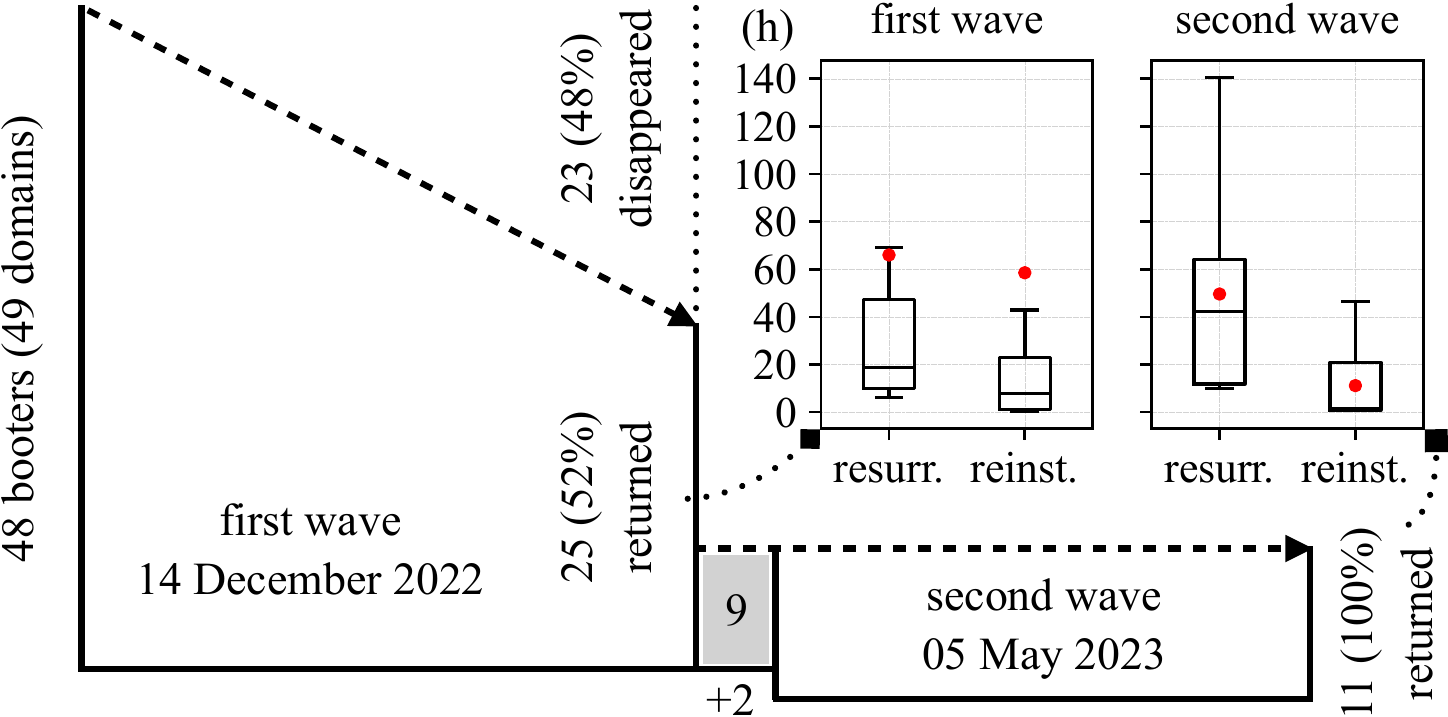}\\
    \caption{Overview of booter resurrections and reinstallations after two waves of takedown (hours). Red dots indicate means.}
    \label{fig:takedown-overview}
\end{figure}
\para{Chat Channels} Many booters operate chat channels to advertise successful attacks, deliver updates, and assist users. Some do further vetting such as click-to-join tasks, but messages are generally public. We monitor Telegram channels of working booters to track resurrections (if any) by capturing new domains being announced. This data is also used to explore how users react to takedowns. As our registered accounts of booters received no email about new domains upon takedowns, the information provided in these channels is likely reliable as booters are incentivised to keep customers quickly informed. 

We collected \BTnTelegramMessages~messages, \BTnTelegramReplies~replies, and associated metadata including \BTnTelegramEmojis~emoji reactions in \BTnCollectedTelegramChannels~channels (both chats and news) from 1 October 2022 to 30 September 2023. As booters may periodically delete historical data, our scraper runs in near real-time using Telethon (with official Telegram APIs), ensuring completeness. Booters may use Discord; we also monitored these channels, which were highly volatile, incomplete, inactive, and often banned rapidly by Discord. They attract far fewer subscribers than Telegram channels, making Telegram the likely primary communication platform.

\section{Takedowns and Resurrections} \label{sec:takedown-resurrection}
This section discusses the disruptive effects on the provision of booters. The domains appeared to have all been taken down on the same day, but there might be a slight lag between the seizure time and when the traffic was redirected to our splash pages. We consider the first wave and second wave to be as of mid-day on 14 December 2022 and 5 May 2023, respectively.

\subsection{The Resurrections} \label{subsec:booter-resurrections}
Booter operators exhibited varying responses upon seizures: some gave up, some waited for a few days to assess the situation, and some promised to return but never did. However, many attempted to recover promptly (after erasing logs for safety reasons), requiring new account registrations, and compensating users for downtime with subscription extensions. Some operators maintain a pool of purchased domains for fast switching, but many acquired new domains only after seizure. Users were informed shortly after new domain purchases, but there were often delays for the sites to return online.

We measure \textit{resurrection} time from seizure to the first successful return, and \textit{reinstallation} time from domain registration to the first successful return (thus, is a subset of \textit{resurrection}). WhoisXML API is used to track domain registrations;\footnote{~WhoisXML Domain \& IP Data Intelligence (https://whoisxmlapi.com/). An alternative way is to use Certificate Transparency, however not all booter domains have certificates issued by authorities providing logs, and the certificate issuance time may not necessarily align with domain registration.} we count the newest among (multiple) historical records returned. We only analyse newly created domains, excluding seven domains in the 1\textsuperscript{st} wave and one in the 2\textsuperscript{nd} wave which were actually old ones being reused, with historical records going back years. \autoref{fig:takedown-overview} overviews the takedowns, \textit{resurrections} and \textit{reinstallations} of seized booters. 

The 1\textsuperscript{st} wave seized \BTnFirstWaveSeizedBooters~booters (\BTnFirstWaveSeizedDomains~domains: \BTnFirstWaveUSRegistrar~(\BTnFirstWaveUSRegistrarProps\%) were registered in the US and \BTnFirstWaveNameCheapRegistrar~(\BTnFirstWaveNameCheapRegistrarProps\%) by Namecheap).\footnote{~Self-attack testing did not work with one booter, and another had both hosting and domains located outside the US, thus were left to be seized by local law enforcement.} Among seized booters, \BTnFirstWaveGaveUpBooters~(48\%) did not return, while \BTnFirstWaveResurrectedBooters~(52\%) resurrected in subsequent days (median \BTnMedianFirstWaveResurrection~hours, mean \BTnMeanFirstWaveResurrection~hours). Of the resurrected domains, 19 (76\%) closely resembled old ones, with top-level domains substituted while second-level domains remained mostly unchanged. The median \textit{installation} time is \BTnMedianFirstWaveReinstallation~hours (mean \BTnMeanFirstWaveReinstallation~hours). 

\begin{figure}[t]
    \centering
    \includegraphics[width=0.475\textwidth]{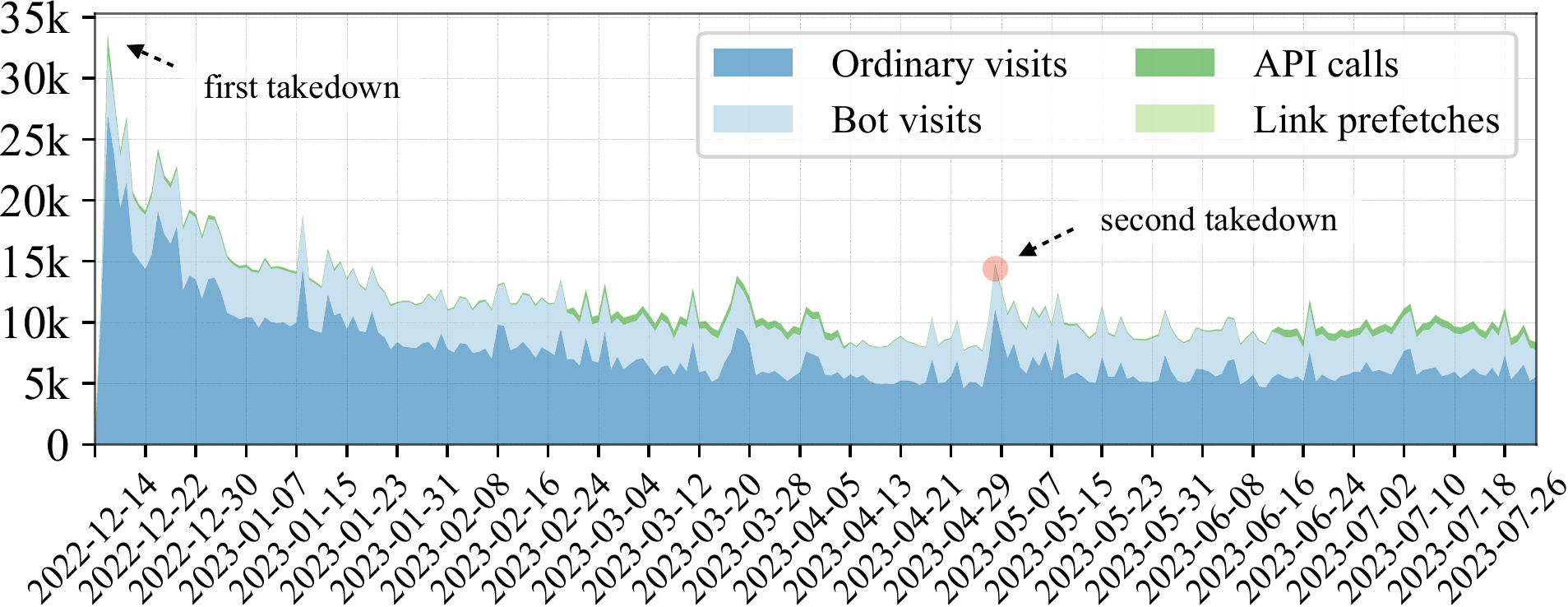}\\
    \caption{The aggregated ground-truth visit sessions per day to booter domains during both waves seen by our splash pages.}
    \label{fig:splash-page-traffic-overview}
\end{figure}
The 2\textsuperscript{nd} wave seized \BTnSecondWaveSeizedBooters~booters: nine were previously seized and two were new. Among the \BTnSecondWaveSeizedDomains~seized domains, \BTnSecondWaveUSRegistrar~(\BTnSecondWaveUSRegistrarProps\%) were registered in the US and \BTnSecondWaveNameCheapRegistrar~(\BTnSecondWaveNameCheapRegistrarProps\%) by Namecheap, showing a decline of reliance on US-based entities. All 11 seized booters this time reappeared under new domains closely resembling old ones, with the \textit{resurrection} median of \BTnMedianSecondWaveResurrection~hours (mean \BTnMeanSecondWaveResurrection~hours, some were around just 1 hour) and \textit{reinstallation} median of \BTnMedianSecondWaveReinstallation~hours (mean \BTnMeanSecondWaveReinstallation~hours).

Regarding the \BTnSecondWaveResurrectedTwiceBooters~booters seized in both waves (i.e. were seized, resurrected, but were seized again), the \textit{resurrection} median is \BTnMedianResurrectionTwiceCombinedBothWave~hour (mean is \BTnMeanResurrectionTwiceCombinedBothWave~hours), while the \textit{reinstallation} median is just \BTnMedianReinstallationTwiceCombinedBothWave~hour (mean is \BTnMeanReinstallationTwiceCombinedBothWave~hours). The shorter \textit{reinstallation} time indicates operators become quicker at purchasing domains and making them available online, presumably as they were more prepared in the second takedown.\footnote{~Due to the inherently unavoidable small sample sizes, we are unable to perform reliable statistical tests comparing means or medians of resurrection/reinstallation between two waves and between US vs. non-US registrars.}

\para{Takeaways} Many of the seized booters in both waves of takedowns resurrected quickly, some within a few hours.

\subsection{Ground-Truth Insights} \label{sec:ground-truth-insights}
Our ground-truth visits include traffic to both seized and NCA-deployed domains. During the first two weeks after the 1\textsuperscript{st} wave, visits to seized domains dropped sharply from over 30k to around 15k per day, see~\autoref{fig:splash-page-traffic-overview}. These declined gradually before returning to about 15k after the 2\textsuperscript{nd} wave, as more domains seized led to more visits recorded and users moving across. However, this temporary increase lasted for just a few days. The proportion of bot visits accounted for around 15\% of traffic at the beginning, but constituted 40-50\% by the end of the period. API calls accounted for only \BTnapicallVisitsProps\%, mostly going to the big booters with some increases around March and July 2023. Our analyses exclude bots and link prefetches.

\begin{figure}[t]
    \centering
    \includegraphics[width=0.4655\textwidth]{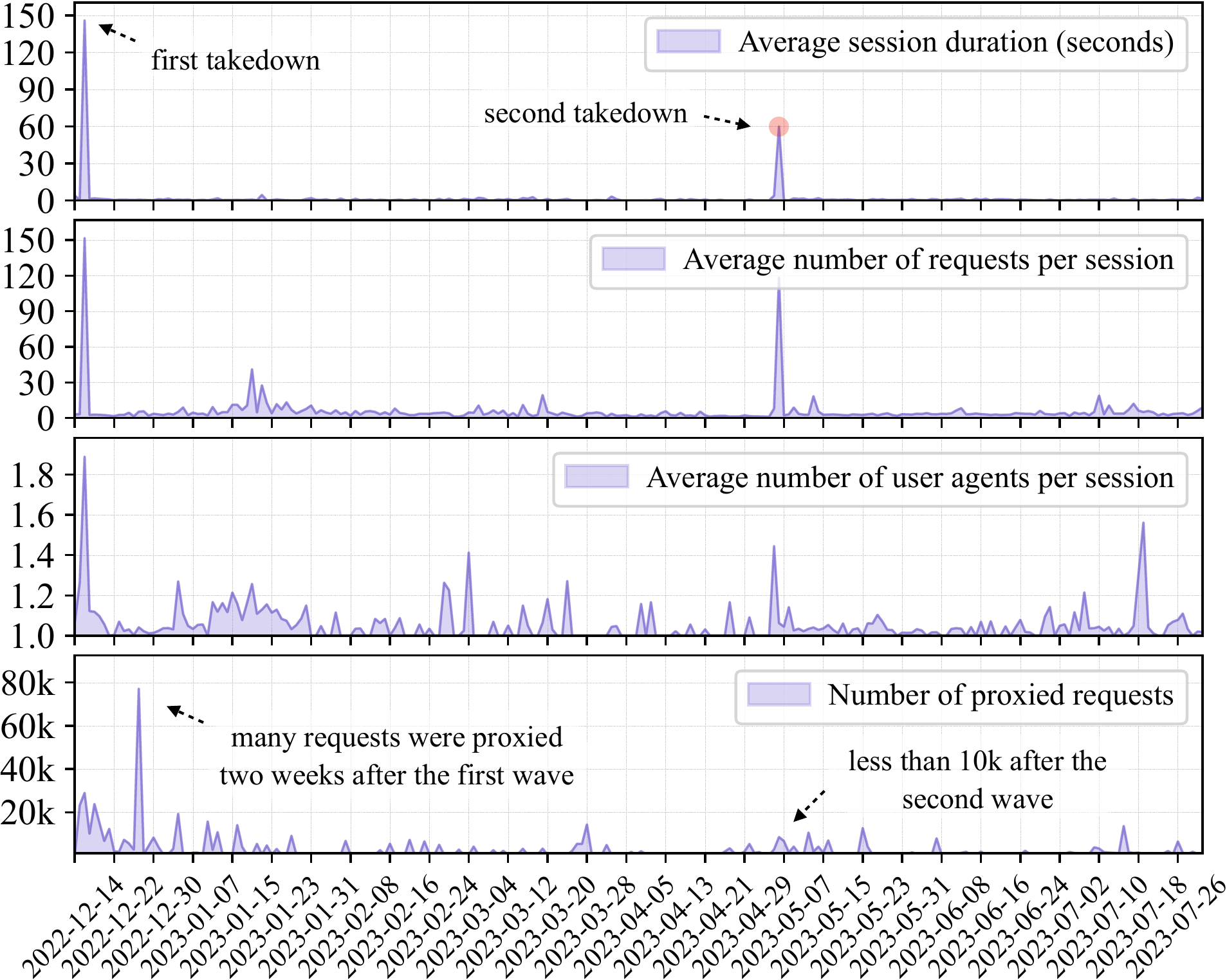}\\
    \caption{The average duration of ground-truth ordinary visit sessions, number of requests and user agents per session, and number of requests made through public proxies and VPNs.}
    \label{fig:splash-page-insights}
\end{figure}
\para{Access Information} Most sessions were short-lived, lasting a few seconds, but were much longer at around 150 and 60 seconds after the 1\textsuperscript{st} and 2\textsuperscript{nd} waves occurred (see \autoref{fig:splash-page-insights}). Among \BTnrealvisitVisits~ordinary sessions, \BTnGroundTruthSessionsWithSingleRequest~(\BTnGroundTruthSessionsWithSingleRequestProps\%) had only 1 initial request (users came and left immediately), while \BTnGroundTruthSessionsWithMultipleRequests~(\BTnGroundTruthSessionsWithMultipleRequestsProps\%) had subsequent requests. The average number of requests per session was 150 and 120 in the 1\textsuperscript{st} and 2\textsuperscript{nd} waves, respectively. The majority of sessions were with a consistent agent, but many got switched often on the takedown days (1.9 and 1.4 on the 1\textsuperscript{st} and 2\textsuperscript{nd} waves), suggesting that users change browsers or fake agents to attempt re-accessing seized sites after noticing the takedowns. A spike occurred on 18 July 2023 (1.6 agents), without a clear reason.

Visits were centralised around top seized domains and access geolocation. The top 5 domains attracted \BTnGroundTruthTopDomainsProps\% visits: the biggest one accounted for \BTnGroundTruthTopDomainsipstresserxcomProps\%, the second took \BTnGroundTruthTopDomainsinstantstresserxcomProps\%, and the next are \BTnGroundTruthTopDomainsstresserxappProps\%, \BTnGroundTruthTopDomainsbootyouxnetProps\%, and \BTnGroundTruthTopDomainsfreestresserxsoProps\%. The top 10 countries accounted for \BTnGroundTruthTopCountriesProps\% visits, where US visitors took \BTnGroundTruthTopCountriesUSProps\%, then China (\BTnGroundTruthTopCountriesCNProps\%), Germany (\BTnGroundTruthTopCountriesDEProps\%), the UK (\BTnGroundTruthTopCountriesGBProps\%), Russia (\BTnGroundTruthTopCountriesRUProps\%); France, the Netherlands, Turkey, Poland, and Singapore took around 2-4\% each.

The browsers and operating systems used suggest a distinctive pattern of technology use compared to the global average. The most frequently used browsers were Chrome, which has a global user market share of 65\% (in 2023) but accounted for only \BTnGroundTruthTopBrowserChromeProps\% of our users, and Firefox (global user market share of 3\% and \BTnGroundTruthTopBrowserFirefoxProps\% of our users). Safari, Chrome Mobile, and Edge each accounted for around 5-7\% of our users (Safari has a global user market share of 18\%). Regarding operating systems, Windows accounted for over \BTnGroundTruthTopOSWindowsProps\% (compared to its 72\% share of the global desktop market) followed by Mac OS (\BTnGroundTruthTopOSMacOSXProps\%), with the remainder Android, iOS, and Linux accounting for 4-10\% each. Over \BTnGroundTruthTopDeviceTypePCProps\% of visits were from PCs, while \BTnGroundTruthTopDeviceTypeMobileProps\% came from mobile devices and only \BTnGroundTruthTopDeviceTypeTabletProps\% from tablets. This suggests a possibly surprising dominance of PC users, given the presumed young user base with a cultural commitment to alternative technology and desktop gaming.

\begin{figure}[t]
    \centering
    \includegraphics[width=0.45\textwidth]{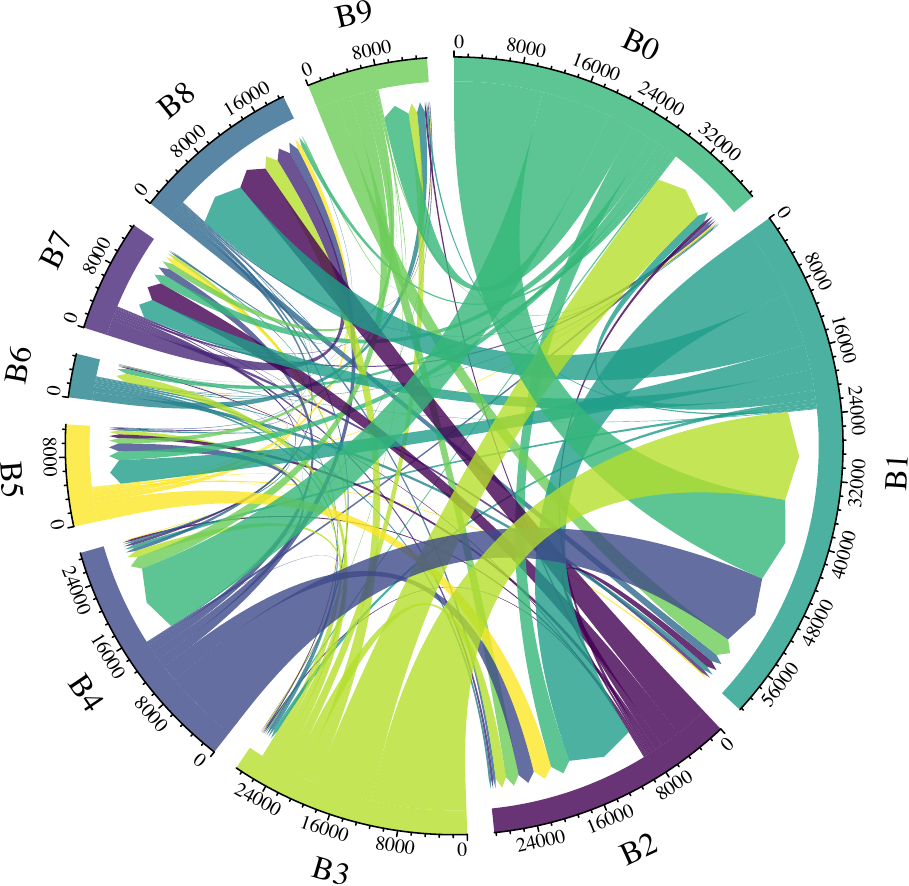}\\
    \caption{The flow of users visiting another seized domain after accessing a seized one, excluding self-navigations. B0 to B9 mark the seized booters that are most frequently navigated.}
    \label{fig:booter-navigations}
\end{figure}
Users may obfuscate IP addresses to hide their identity, especially for cybercrime services. To detect proxy usage, we used IP2Proxy, which covers various types (e.g., VPNs, open proxies, residential proxies, Tor exit nodes). Unlike in the crypto mixing market, where over 60\% of transactions are made through anonymous tunnels~\cite{miedema2023mixed}, users accessing seized booters generally did not use proxies. At the session level, \BTGroundTruthSessionsUsingProxy~of a total of \BTnTotalVisitSessions~visit sessions (\BTGroundTruthSessionsUsingProxyProps\%) used public proxies on first access, with \BTGroundTruthRequestsUsingProxy~(\BTGroundTruthRequestsUsingProxyProps\%)~using proxies from a total of \BTnTotalRawEvent~raw requests. The number of proxied requests peaked at around 80k on 27 December 2022 (two weeks after the 1\textsuperscript{st} wave) then declined gradually, but only around 10k proxied requests were found after the 2\textsuperscript{nd} wave, see \autoref{fig:splash-page-insights}. Over \BTGroundTruthRequestsUsageTypeDCHProps\% of proxied requests were from data centres and web hostings, \BTGroundTruthRequestsUsageTypeISPProps\% were from fixed line ISPs, while the other sources are trivial. Despite its fairly widespread media association with crime, no requests were made through Tor, aligning with a relatively technically-unskilled user base with no real adoption of basic operational security practices. 

\begin{figure}[t]
    \centering
    \includegraphics[width=0.475\textwidth]{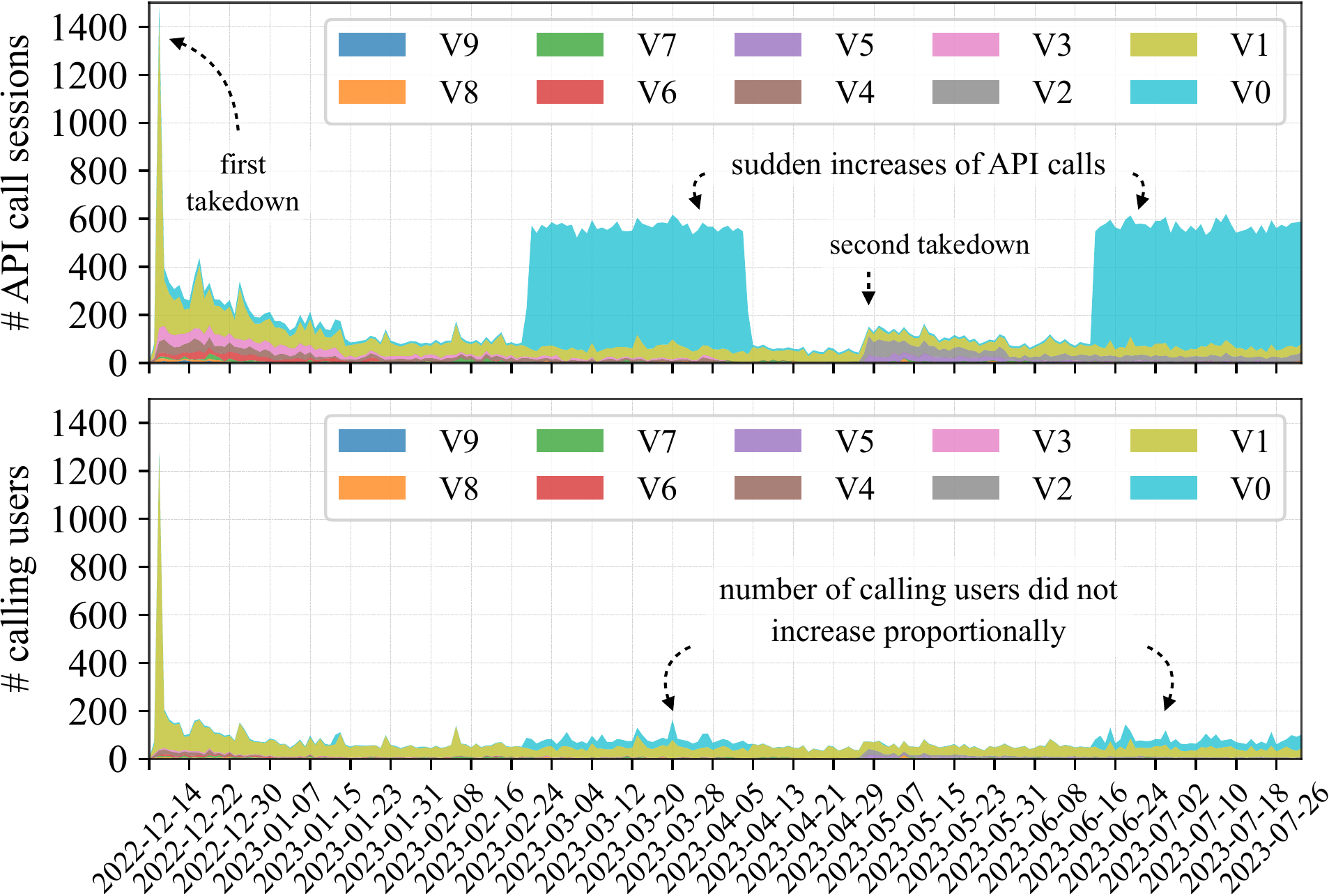}\\
    \caption{Number of API request sessions (top) and calling users (bottom) per day by top booter vendors over the period.}
    \label{fig:reselling-capacity}
\end{figure}
\begin{figure*}[t]
    \centering
    \includegraphics[width=1\textwidth]{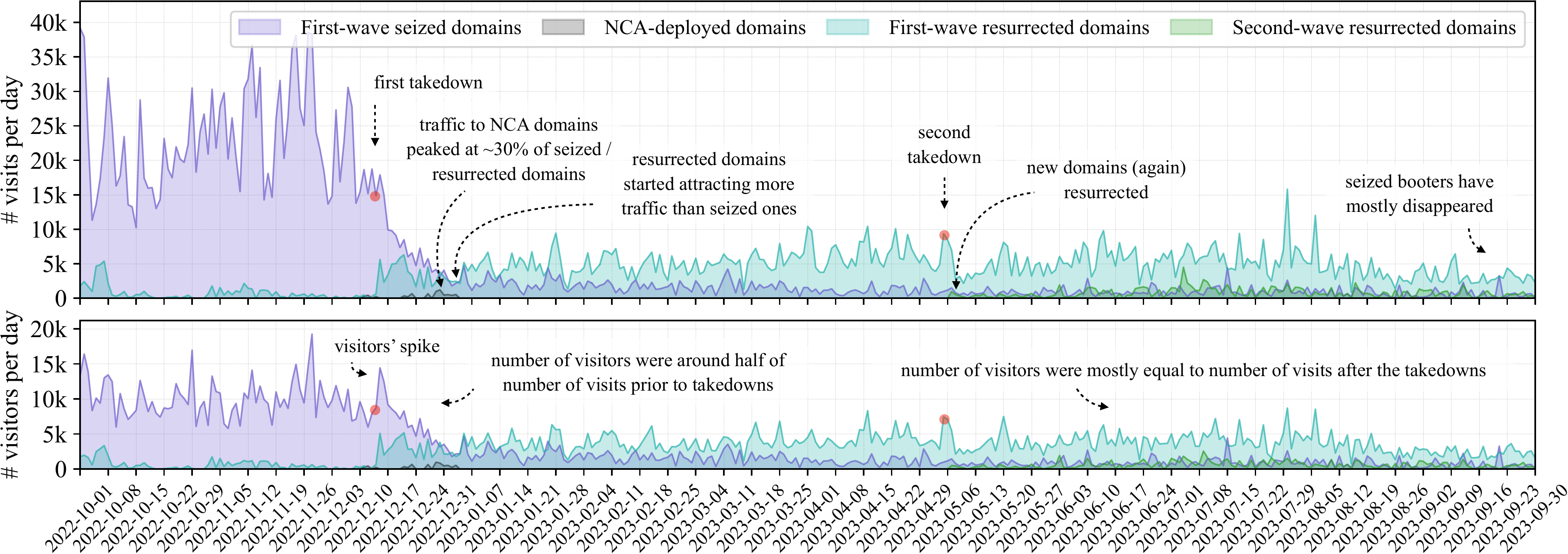}\\
    \caption{Number of web visits and visitors to seized domains (\BTnFirstWaveSeizedDomainsThatHasSimilarwebTraffic~first-wave, \BTnSecondWaveSeizedDomainsThatHasSimilarwebTraffic~second-wave), resurrected domains (\BTnFirstWaveResurrectedDomainsThatHasSimilarwebTraffic~first-wave, \BTnSecondWaveResurrectedDomainsThatHasSimilarwebTraffic~second-wave), and all NCA deceptive domains per day. Before the takedowns, there were some visits to first-wave resurrected domains as they were pre-purchased then reused. Most of the second-wave seized domains were first-wave resurrected domains.}
    \label{fig:booters-traffic}
\end{figure*}
\begin{figure}[t]
    \centering
    \includegraphics[width=0.475\textwidth]{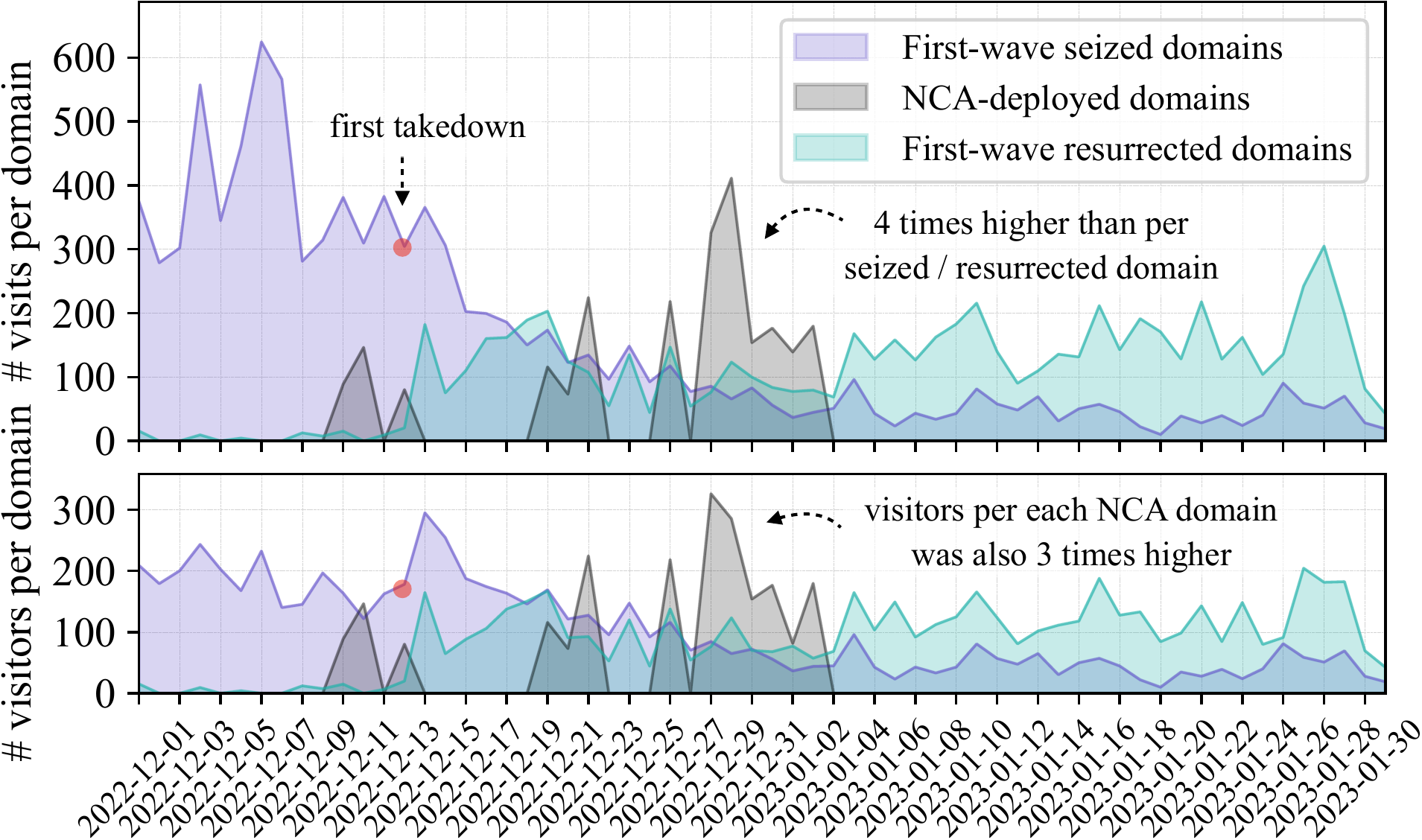}\\
    \caption{Number of web visits and visitors per each first-wave seized and resurrected domain, as well as NCA domain.}
    \label{fig:booter-traffic-nca}
\end{figure}
\para{Navigation Across Booters} Users may seek and navigate to other booters when one is disrupted, but their methods to find new services are not well understood. A major booter dashboard deliberately shut down after the 1\textsuperscript{st} wave, limiting options to find new ones if users were not in relevant channels, did not have prior knowledge, or did not follow informative sources. Booters disappeared quickly from search engines, with several government removal requests~\cite{ncalumendatabase}; the first few pages in the search results for DDoS and booting are mostly about news, takedowns, and legal implications. Among \BTNIPAccessingDomainsTotal~unique accessing IP addresses, \BTNIPAccessingOnlyOneDomainProps\% visited just one domain while \BTNIPAccessingMultipleDomainProps\% visited multiple ones (among those \BTNIPAccessingMoreThanFiveDomainProps\% visited more than five and \BTNIPAccessingMoreThanTenDomainProps\% visited more than 10), suggesting at least around one fourth of users had pre-existing awareness of multiple booters, and some know many. 

We define a \textit{navigation} from a seized booter $b_1 \rightarrow b_2$ as when users visit $b_2$ right after $b_1$; thus, $b_1$ makes an \textit{outbound} to $b_2$, and $b_2$ has an \textit{inbound} from $b_1$. The most frequently navigated booters are those with the most inbounds. Users accessing top seized booters tend not to navigate to the others: among the top 10 most navigated booters, only B2 is among the 10 largest booters prior to takedown, see~\autoref{fig:booter-navigations}. The five most navigated booters have strong ties. The biggest navigating booter was B0 with \BTNavigationCountsFromipstressercom~outbounds mostly to B1 (\BTNavigationCountsFromipstressercomToinstantstressercom) and B4 (\BTNavigationCountsFromipstressercomTofreestresserso). The most navigated booter was B1 with \BTNavigationCountsToinstantstressercom~inbounds, mainly from B3 (\BTNavigationCountsFrombootyounetToinstantstressercom), B0 (\BTNavigationCountsFromipstressercomToinstantstressercom), and B4 (\BTNavigationCountsFromfreestressersoToinstantstressercom). However, there was mostly no outbounds from B1 back to those three domains, but to B8 (\BTNavigationCountsFrominstantstressercomTostresseraicom) and B2 (\BTNavigationCountsFrominstantstressercomTostresserapp). The other major ones (B2, B3, B4) also have different major inbounds and outbounds, suggesting their relationships are mostly unidirectional despite having strong ties.

The NCA's deceptive domains attracted \BTNavigationCountsToNCADomains~inbounds from and \BTNavigationCountsFromNCADomains~outbounds to seized booters, while there were \BTNavigationCountsWithinAllNCADomains~navigations between NCA domains. These are not shown in \autoref{fig:booter-navigations}~as NCA domains are not among the top 10.

\para{Reselling Capacity} Booters may operate as intermediate services, using APIs provided by others to create second-tier booters e.g., Webstresser resold attack capacity to smaller ones before being seized~\cite{collier2020cybercrime}. Our ground-truth data reveals booters providing the APIs, but not who calls the APIs as it is not reliably feasible to match IP sources with booter identities (we do not know the IP addresses of their back-end services). 

The patterns of API call sessions and calling users generally mirror the visit traffic. There was a sharp decline from around 1\,400 to less than 300 API call sessions per day within a week after the 1\textsuperscript{st} wave (a decline of around 80\%), see \autoref{fig:reselling-capacity}. The number of calling users plummeted even more dramatically, from over 1\,200 to around 100 per day (a decline of over 90\%). The 2\textsuperscript{nd} wave coincided with an increase in API calls from around 50 to 150 per day, mainly due to more domains being seized; however, this effect waned quickly. Among the total \BTnapicallVisits~sessions, the top 10 booter providers (vendors) provided \BTNApiCallSessionsTopTenProps\%, accessed by \BTNApiCallingUsersTopTen~unique IP addresses (\BTNApiCallingUsersTopTenProps\%~of the total accessing IPs). Four of these top 10 vendors also rank among the 10 largest booters (1\textsuperscript{st}, 4\textsuperscript{th}, 6\textsuperscript{th}, and 7\textsuperscript{th}).

The largest vendor V1 initially dominated the market but was surpassed in March 2024 by V0 with a sudden surge of around 500 per day, lasting for 1.5 months before returning to near zero then raising again from late June to the end of July 2023. These increases were mainly attributed to a single user (the same API key) repeatedly requesting 180-minute TCP-based attacks to target a few victims under ports 80, 443, and 2222. This was likely a premium subscriber making multiple requests due to unsuccessful attempts, suggesting that users might not notice the interventions and their backends continued calling APIs months afterwards. We lack evidence to determine if that is an individual or a vendor, as any premium licensees can run simultaneous attacks, thus reselling capacity. Given the scale and that normal users would likely verify the success of attacks, we lean toward the likelihood of a reseller. V2 and V3 emerged after the 2\textsuperscript{nd} wave, but their contributions were short-lived. Other vendors followed a consistent pattern, gradually declining to around 50 per day by the end of July.

\para{Takeaways} Users were aware of multiple booters and moved between them after the seizures. Some did not notice the takedown and continued making hundreds of API requests.

\section{The Longitudinal Effects} \label{sec:takedown-effects}
We now discuss the longitudinal effects on traffic to booter domains, DDoS attack counts, and community perspectives.

\subsection{Traffic to Booter Domains} \label{subsec:traffic-to-booter-domains}
Section~\ref{sec:ground-truth-insights} shows a clear decline of ground-truth visits to seized domains after the takedowns. Our Similarweb data further reveals traffic to resurrected domains and NCA domains.

\begin{figure}[t]
    \centering
    \includegraphics[width=0.475\textwidth]{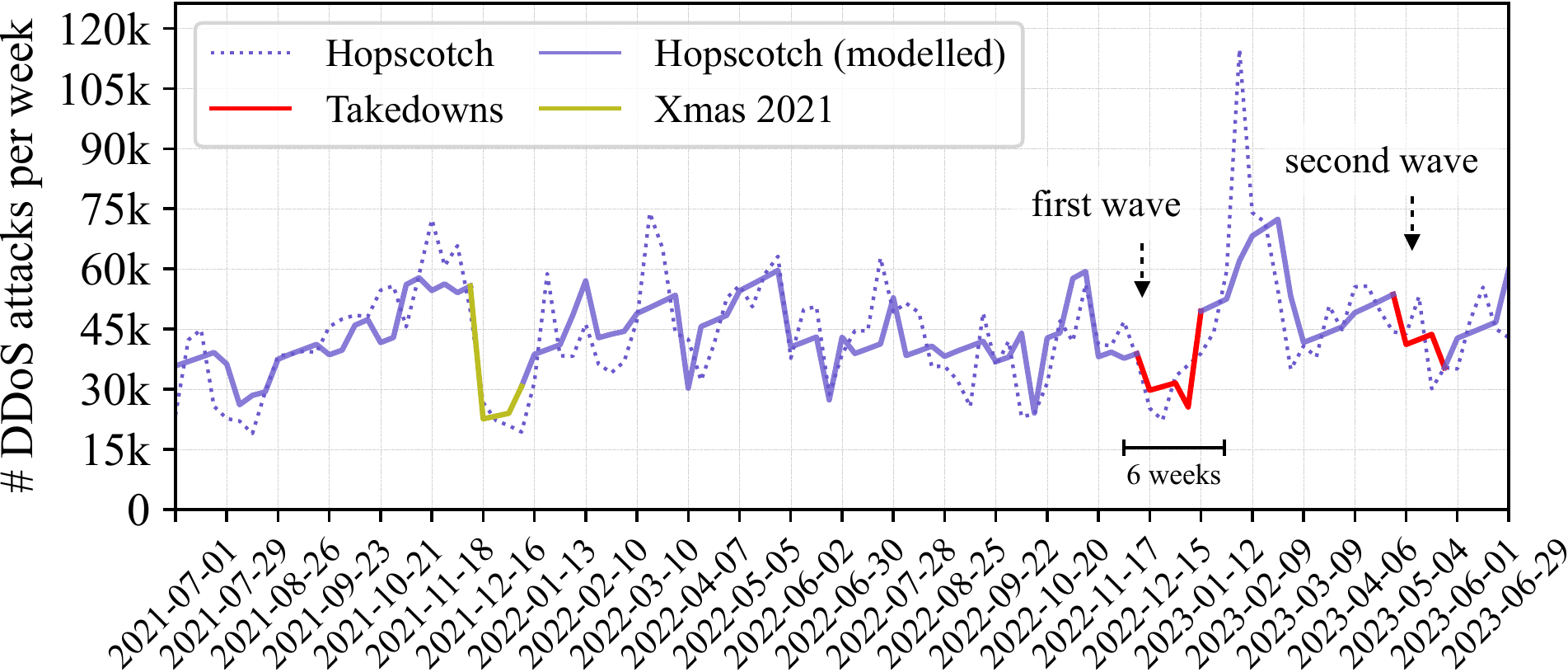}\\
    \caption{Modelled weekly DDoS attacks (\hopscotch)}
    \label{fig:ddos-hopscotch}
\end{figure}
\begin{table}[t]
\centering
\vspace{-4mm}
\caption{The negative binomial regression model composition with takedowns and seasonal components $S_i$ (\hopscotch).}
\setlength{\tabcolsep}{0.225em}
\vspace{2mm}
\small
\begin{tabular}{lccccccc}
    \toprule
    Type & Events & Coef. & Std. Err. & $z$ & $P>|z|$ & \multicolumn{2}{c}{[95\% CI]}\\
    \midrule
    \multirow{15}{*}{UDP} & 1\textsuperscript{st} wave & -0.4991 &  0.241 & -2.069 &  * &  -0.972 &  -0.026\\
    & 2\textsuperscript{nd} wave & 0.2402 &  0.281 &  0.854 &  0.393 &  -0.311 &   0.792\\
    & Xmas'21 & -0.4587 &  0.213 & -2.158 &  * &  -0.875 &  -0.042\\
    & $S_{1}$ &  -0.5373 &  0.250 & -2.146 &  * &  -1.028 &  -0.047\\
    & $S_{2}$ &  -0.4993 &  0.259 & -1.931 &  0.053 &  -1.006 &   0.007\\
    & $S_{3}$ &  -0.1851 &  0.255 & -0.726 &  0.468 &  -0.685 &   0.314\\
    & $S_{4}$ &  -0.1817 &  0.261 & -0.695 &  0.487 &  -0.694 &   0.330\\
    & $S_{5}$ &  -0.2421 &  0.260 & -0.933 &  0.351 &  -0.751 &   0.267\\
    & $S_{6}$ &  -0.2303 &  0.257 & -0.895 &  0.371 &  -0.735 &   0.274\\
    & $S_{7}$ &  -0.6059 &  0.257 & -2.354 &  * &  -1.110 &  -0.101\\
    & $S_{8}$ &  -0.3060 &  0.252 & -1.212 &  0.225 &  -0.801 &   0.189\\
    & $S_{9}$ &  -0.0759 &  0.221 & -0.343 &  0.732 &  -0.510 &   0.358\\
    & $S_{10}$ &  0.0881 &  0.205 &  0.430 &  0.667 &  -0.313 &   0.490\\
    & $S_{11}$ & -0.0873 &  0.157 & -0.555 &  0.579 &  -0.395 &   0.221\\
    \bottomrule
\end{tabular}
\\{\vspace{1mm}\hspace{2mm}\raggedright 
\footnotesize{Level of significance: * ($p < 0.05$), ** ($p < 0.01$), *** ($p < 0.001$).} \par}
\label{tab:nbr-composition-hopscotch}
\end{table}
\para{Traffic Displacement} The 1\textsuperscript{st} wave corresponded with a reduction in visits to the seized domains by 80-90\%, from around 20-25k per day to 2.5k after two weeks and to near zero after one year (see~\autoref{fig:booters-traffic}). There was also a rapid decline of over 75\% of visitors after two weeks (almost 100\% after one year), with a sudden but short-lived increase of 50\% right after the 1\textsuperscript{st} wave as users visited multiple seized booters (see more in~\S\ref{sec:ground-truth-insights}). The pre-takedown ratio of visitors per visit was around 1:2 (meaning users visited a domain twice per day), but dropped to almost 1:1 after the 1\textsuperscript{st} wave, suggesting that users left quickly after visiting once.

Traffic was fragmented to first-wave resurrected domains, which grew quickly and stabilised at roughly 5k per day after two weeks, outpacing traffic to seized domains. However, this is still around 80-90\% less than the prior traffic to the seized domains. The 2\textsuperscript{nd} wave associated with a further decrease in traffic to first-wave resurrected domains by 70\%, but this time traffic quickly recovered after 2 weeks, showing a mild effect. This suggests the persistence of a smaller and more committed user base post the first wave. Booters attempted to resurrect again, but traffic to the second-wave resurrected domains was trivial, declining gradually to almost zero. At the end of the period, all domains combined only attracted around 4k visits per day, presumably due to domain expiration, users avoiding repeatedly accessing seized domains, or users not noticing new domains in chat channels; but overall the resurrections have failed to attract as many visits and visitors as before.

\para{Traffic to NCA-Deployed Domains} The NCA's UK-focused influence campaign attracted \BTnTotalNCATrafficVisitsDuringTwoMonthsWeb~web visits to the `fake' domains during two months around the 1\textsuperscript{st} wave, peaking on 30 December 2022 at \BTnTotalNCATrafficVisitsDuringTwoMonthsPeakWeb~(around one-third of visits to first-wave seized and resurrected domains on that day, see~\autoref{fig:booters-traffic}). The number of visits and visitors per each domain were four and three times higher than those of first-wave seized and resurrected domains (see~\autoref{fig:booter-traffic-nca}). However, the peaks dropped quickly and these domains became mostly inactive after about one week, suggesting a notable but brief effect of the NCA's information penetration to confuse the UK market.

\para{Takeaways} Re-emerged booter domains, despite resembling old ones, did not attract the same amount of traffic. The NCA-deployed booters attracted some visitors, but they left quickly.

\subsection{DDoS Attack Volumes} \label{subsec:ddos-attack-volumes}
As other potential events may occur during the period, along with seasonal effects, precisely quantifying and identifying cause-and-effect signals based on raw (noisy) data is difficult. To test the significance of the impact, we modelled the \textit{weekly} attack counts (as daily counts are noisy) from \hopscotch, \amppot, \netscout, and self-report datasets using negative binomial regression -- a well-established statistical technique for interrupted time series analysis that incorporates various components, including the underlying trends, random variation, seasonal variations (which are often responsible for periodic dips), and intervention effects~\cite{harvey1989time}. Intervention models are then incrementally built and tested for fit from the core seasonal-plus-trend linear model, producing a final model that indicates the size, duration, and scope of a range of intervention effects. Having been used widely by researchers~\cite{braga2008policing,steinbach2015effect}, this model suits DDoS attack counts~\cite{collier2019booting}, considering not only whether an absolute reduction in values is observed following an intervention but also the difference from the expected counts based on previous trends and seasonality. 

These models are theory- and domain-knowledge-driven, with researchers selecting intervention components and durations based on prior knowledge and theorised effect models, then testing the resulting models for fit and parsimony. For each takedown, we specify an intervention period through observation and testing of different durations, start, and end points for fit and feasibility, incorporating significant interventions stepwise into an overall model. This allows the effects of previous interventions on the series to be taken into account when assessing the significance of subsequent ones. We also include day-of-week effects, time trend, and seasonal variables that may influence the model.
Applying this to all datasets, we report the model compositions, including the month-by-month seasonality of the data and the (in)significant effects of the two waves, then compare them with the Xmas 2021 event as it occurs in the same season as the 1\textsuperscript{st} wave.

\begin{figure}[t]
    \centering
    \includegraphics[width=0.475\textwidth]{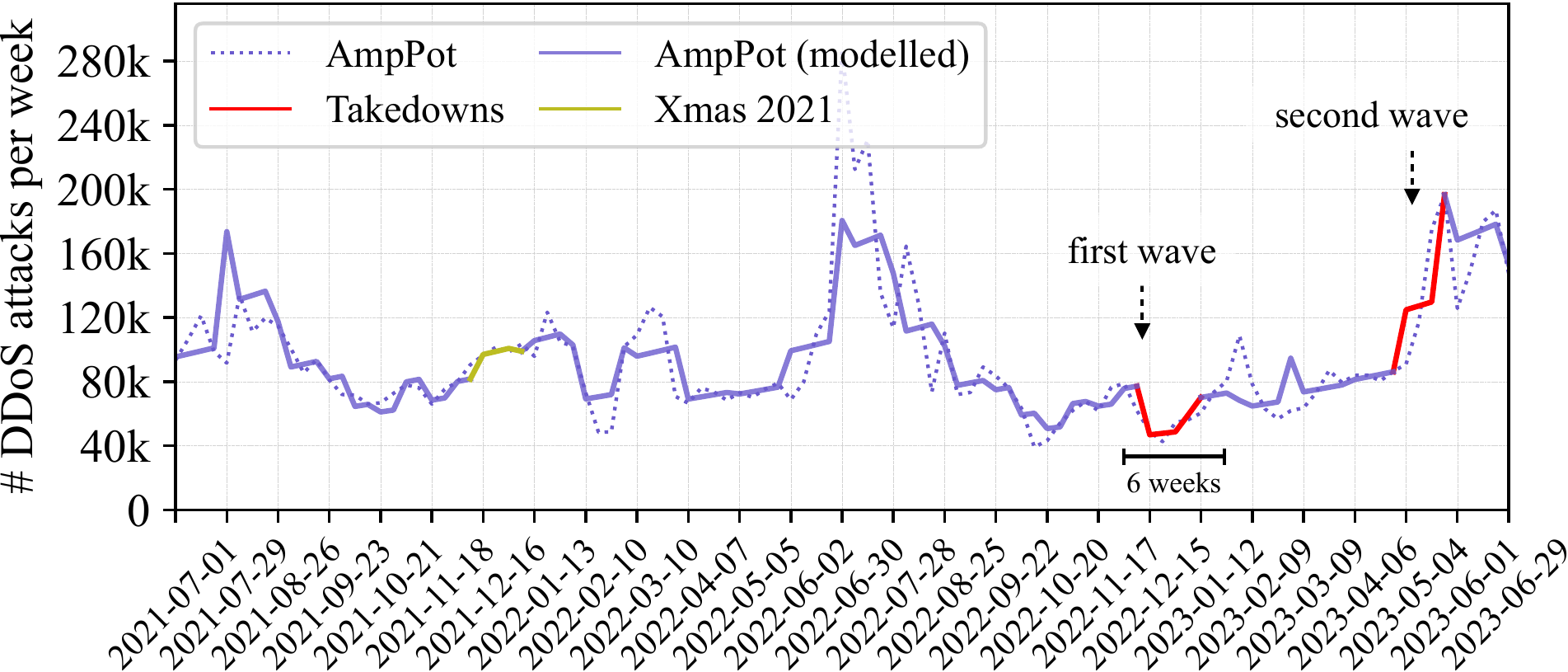}\\
    \caption{Modelled weekly DDoS attacks (\amppot)}
    \label{fig:ddos-amppot}
\end{figure}
\begin{table}[t]
\centering
\vspace{-4mm}
\caption{The negative binomial regression model composition with takedowns and seasonal components $S_i$ (\amppot).}
\setlength{\tabcolsep}{0.225em}
\vspace{2mm}
\small
\begin{tabular}{lccccccc}
    \toprule
    Type & Events & Coef. & Std. Err. & $z$ & $P>|z|$ & \multicolumn{2}{c}{[95\% CI]}\\
    \midrule
    \multirow{15}{*}{UDP} & 1\textsuperscript{st} wave & -0.5642 &  0.218 & -2.584 &  * &  -0.992 &  -0.136\\
    & 2\textsuperscript{nd} wave & -0.3969 &  0.245 & -1.620 &  0.105 &  -0.877 &   0.083\\
    & Xmas'21 & 0.2153 &  0.186 &  1.156 &  0.248 &  -0.150 &   0.580\\
    & $S_{1}$ &   0.9165 &  0.223 &  4.118 &  *** &   0.480 &   1.353\\
    & $S_{2}$ &   1.4183 &  0.231 &  6.131 &  *** &   0.965 &   1.872\\
    & $S_{3}$ &   1.2287 &  0.231 &  5.319 &  *** &   0.776 &   1.681\\
    & $S_{4}$ &   1.0649 &  0.240 &  4.445 &  *** &   0.595 &   1.535\\
    & $S_{5}$ &   0.9513 &  0.240 &  3.956 &  *** &   0.480 &   1.423\\
    & $S_{6}$ &   0.7393 &  0.242 &  3.054 &  ** &   0.265 &   1.214\\
    & $S_{7}$ &   0.6564 &  0.242 &  2.717 &  ** &   0.183 &   1.130\\
    & $S_{8}$ &   0.6825 &  0.240 &  2.839 &  ** &   0.211 &   1.154\\
    & $S_{9}$ &   0.2472 &  0.213 &  1.163 &  0.245 &  -0.169 &   0.664\\
    & $S_{10}$ &  0.1544 &  0.199 &  0.775 &  0.439 &  -0.236 &   0.545\\
    & $S_{11}$ & -0.1361 &  0.153 & -0.889 &  0.374 &  -0.436 &   0.164\\
    \bottomrule
\end{tabular}
\\{\vspace{1mm}\hspace{2mm}\raggedright 
\footnotesize{Level of significance: * ($p < 0.05$), ** ($p < 0.01$), *** ($p < 0.001$).} \par}
\label{tab:nbr-composition-amppot}
\end{table}
\para{The \hopscotch Perspective} The 1\textsuperscript{st} wave coincided with a statistically significant drop in weekly attack counts from around 45k to 25k (see \autoref{fig:ddos-hopscotch} and \autoref{tab:nbr-composition-hopscotch}). This endured for about six weeks until February 2023, when the attack counts rebounded and even surpassed the prior levels. In contrast, the 2\textsuperscript{nd} wave appeared to have unclear and not statistically significant effects. The same interval from the previous year, Xmas'21, also shows a statistically significant decrease.

\begin{figure}[t]
    \centering
    \includegraphics[width=0.475\textwidth]{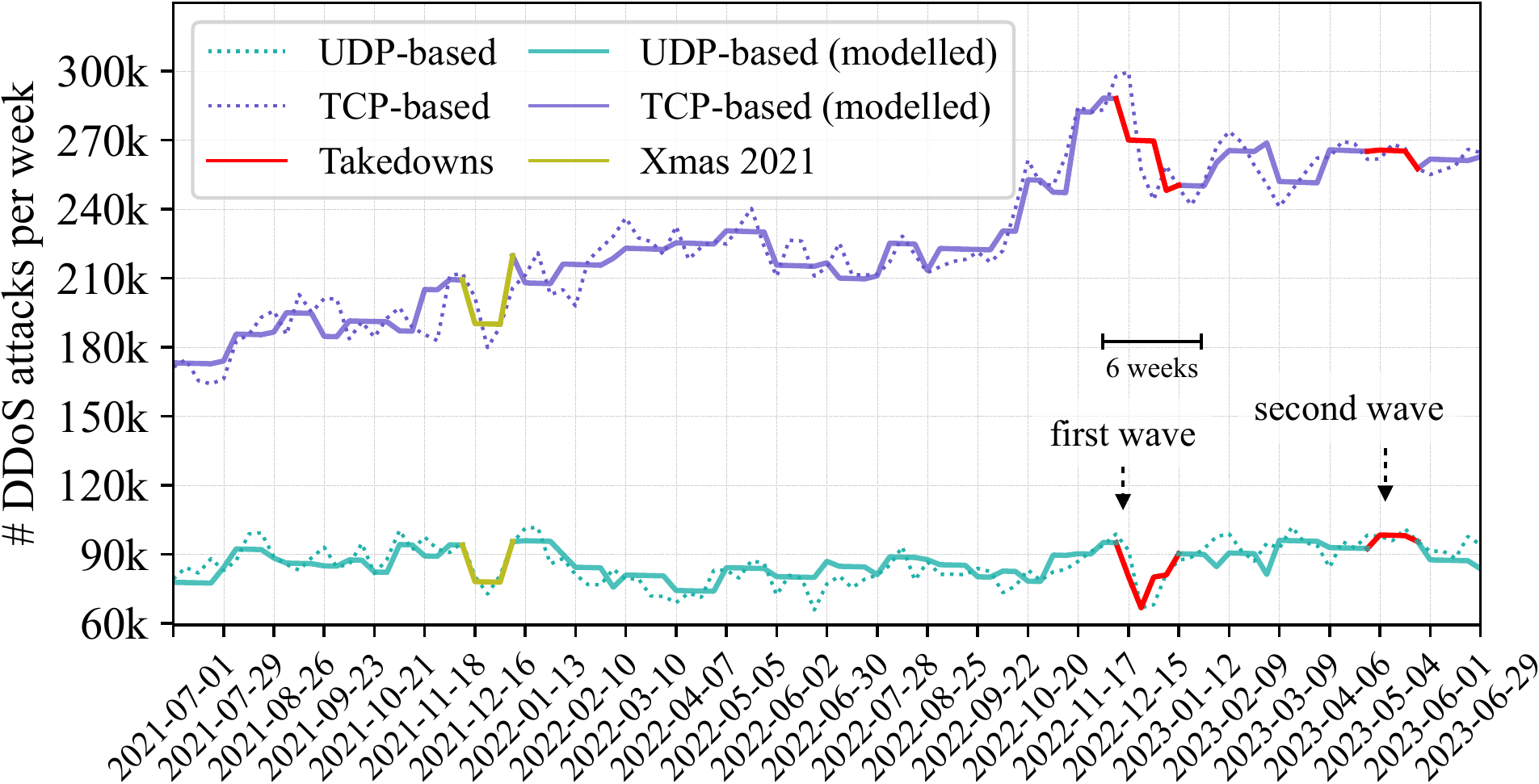}\\
    \caption{Modelled weekly DDoS attacks, as well as those broken down by UDP- and TCP-based attacks (\netscout).}
    \label{fig:ddos-netscout}
\end{figure}
\begin{table}[ht!]
\centering
\vspace{-4mm}
\caption{The negative binomial regression model composition with takedowns and seasonal components $S_i$ (\netscout).}
\setlength{\tabcolsep}{0.225em}
\vspace{2mm}
\small
\begin{tabular}{lccccccc}
    \toprule
    Type & Events & Coef. & Std. Err. & $z$ & $P>|z|$ & \multicolumn{2}{c}{[95\% CI]}\\
    \midrule
    \multirow{15}{*}{TCP} & 1\textsuperscript{st} wave &  -0.0098 &   0.037 &  -0.267 &   0.790 &   -0.082 &    0.062 \\
    & 2\textsuperscript{nd} wave & 0.0283 &   0.037 &   0.757 &   0.449 &   -0.045 &    0.102\\
    & Xmas'21 & -0.2283 &   0.034 &  -6.656 &   *** &   -0.295 &   -0.161\\
    & $S_{1}$ & -0.0725 &   0.036 &  -2.029 &   * &   -0.143 &   -0.002\\
    & $S_{2}$ & -0.0647 &   0.039 &  -1.650 &   0.099 &   -0.142 &    0.012\\
    & $S_{3}$ &  -0.0578 &   0.039 &  -1.469 &   0.142 &   -0.135 &    0.019\\
    & $S_{4}$ & -0.1097 &   0.042 &  -2.613 &   ** &   -0.192 &   -0.027\\
    & $S_{5}$ &  -0.1095 &   0.042 &  -2.625 &   ** &   -0.191 &   -0.028\\
    & $S_{6}$ &  -0.0161 &   0.041 &  -0.393 &   0.695 &   -0.096 &    0.064\\
    & $S_{7}$ &  0.1179 &   0.040 &   2.954 &   ** &    0.040 &    0.196\\
    & $S_{8}$ &  0.0639 &   0.042 &   1.535 &   0.125 &   -0.018 &    0.145\\
    & $S_{9}$ &  0.0638 &   0.035 &   1.843 &   0.065 &   -0.004 &    0.132\\
    & $S_{10}$ &  0.0845 &   0.032 &   2.653 &   ** &    0.022 &    0.147\\
    & $S_{11}$ &  0.0209 &   0.024 &   0.852 &   0.394 &   -0.027 &    0.069\\
    \midrule
    \multirow{15}{*}{UDP} & 1\textsuperscript{st} wave & -0.1722 &   0.064 &  -2.681 &   ** &   -0.298 &   -0.046\\
    & 2\textsuperscript{nd} wave & 0.0248 &   0.068 &   0.366 &   0.714 &   -0.108 &    0.157\\
    & Xmas'21 & -0.1932 &   0.058 &  -3.342 &   ** &   -0.306 &   -0.080\\
    & $S_{1}$ &  0.0906 &   0.064 &   1.416 &   0.157 &   -0.035 &    0.216\\
    & $S_{2}$ &  0.1368 &   0.068 &   2.006 &   * &    0.003 &    0.270\\
    & $S_{3}$ &  0.0583 &   0.069 &   0.849 &   0.396 &   -0.076 &    0.193\\
    & $S_{4}$ &   0.0094 &   0.071 &   0.133 &   0.894 &   -0.130 &    0.149\\
    & $S_{5}$ &  -0.0888 &   0.071 &  -1.245 &   0.213 &   -0.229 &    0.051\\
    & $S_{6}$ &  -0.1793 &   0.072 &  -2.506 &   * &   -0.319 &   -0.039\\
    & $S_{7}$ &   -0.2089 &   0.071 &  -2.958 &   ** &   -0.347 &   -0.070\\
    & $S_{8}$ & -0.2423 &   0.071 &  -3.425 &   ** &   -0.381 &   -0.104\\
    & $S_{9}$ &  -0.3441 &   0.064 &  -5.353 &   *** &   -0.470 &   -0.218\\
    & $S_{10}$ & -0.3139 &   0.060 &  -5.228 &   *** &   -0.432 &   -0.196\\
    & $S_{11}$ & -0.1836 &   0.048 &  -3.826 &   *** &   -0.278 &   -0.090\\
    \bottomrule
\end{tabular}
\\{\vspace{1mm}\hspace{2mm}\raggedright 
\footnotesize{Level of significance: * ($p < 0.05$), ** ($p < 0.01$), *** ($p < 0.001$).} \par}
\label{tab:nbr-composition-netscout}
\end{table}
\para{The \amppot Perspective} There was a decline in weekly attack counts from around 80k to 40k associated with the 1\textsuperscript{st} wave (see \autoref{fig:ddos-amppot}), which is statistically significant (see \autoref{tab:nbr-composition-amppot}). This drop lasted for a similar period as seen by \hopscotch, also followed by a rebound. The 2\textsuperscript{nd} wave effect was not significant, consistent with \hopscotch's view. But contrary to \hopscotch, the Xmas'21 effect here is not statistically significant, with attack counts even slightly increasing.

\para{The \netscout Perspective} There was a statistically significant drop in weekly UDP-based attacks after the 1\textsuperscript{st} wave, from roughly 100k to 70k (see~\autoref{fig:ddos-netscout} and~\autoref{tab:nbr-composition-netscout}). However, attack counts rebounded after a few weeks. TCP-based attacks show a similar decline, from around 300k to 240k, but this was \textit{not} statistically significant over the entire series, suggesting the effect was mostly on UDP-based attacks. The 2\textsuperscript{nd} wave did not have any clear effects on either type of attacks, but Xmas'21 correlated with significant drops in both types.

\begin{figure}[t]
    \centering
    \includegraphics[width=0.475\textwidth]{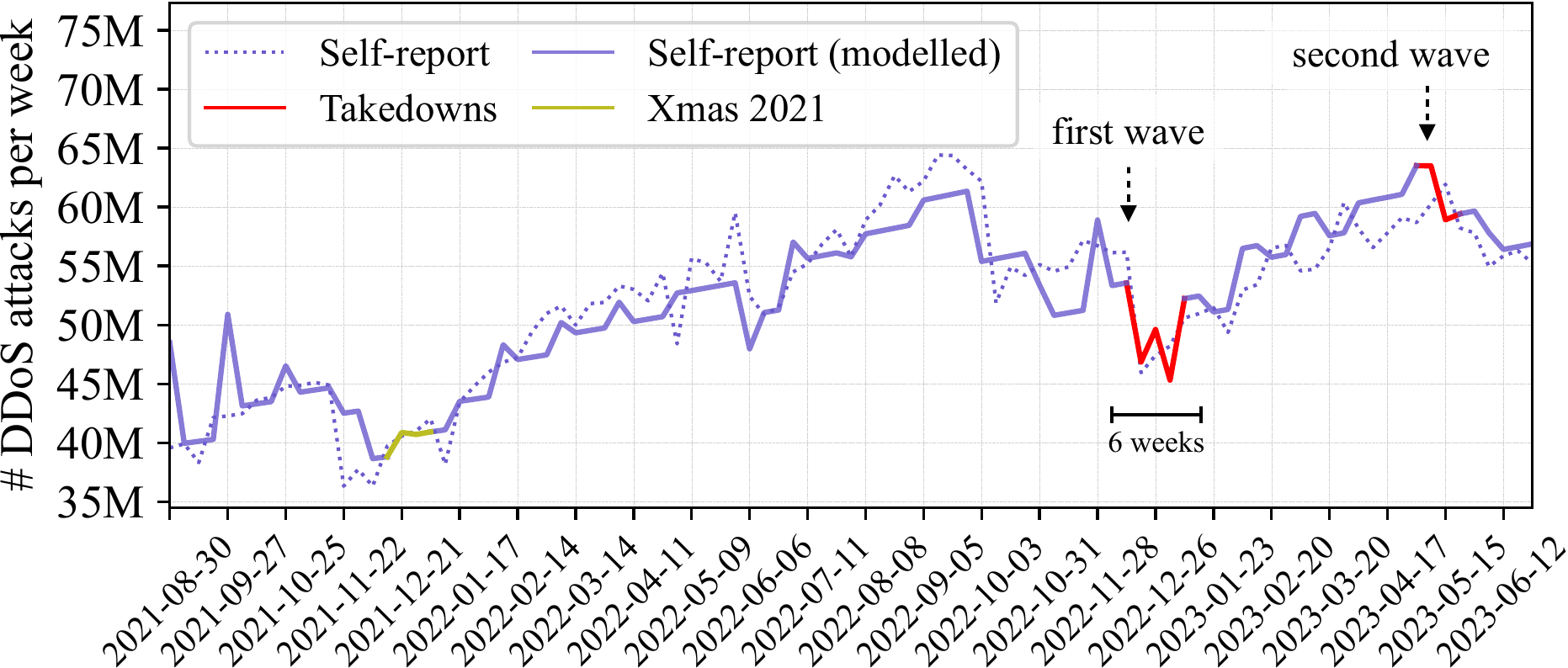}\\
    \caption{Modelled weekly DDoS attacks (self-reported)}
    \label{fig:self-reported-ddos-attacks}
\end{figure}
\begin{table}[t]
\centering
\vspace{-4mm}
\caption{The negative binomial regression model composition with takedowns and seasonal components $S_i$ (self-reported).}
\setlength{\tabcolsep}{0.225em}
\vspace{2mm}
\small
\begin{tabular}{lccccccc}
    \toprule
    Type & Events & Coef. & Std. Err. & $z$ & $P>|z|$ & \multicolumn{2}{c}{[95\% CI]}\\
    \midrule
    \multirow{15}{*}{All} & 1\textsuperscript{st} wave & -0.1378 &  0.042 & -3.262 &  ** &  -0.221 &  -0.055\\
    & 2\textsuperscript{nd} wave & -0.0043 &  0.046 & -0.095 &  0.924 &  -0.094 &   0.085\\
    & Xmas'21 & -0.0960 &  0.051 & -1.865 &  0.062 &  -0.197 &   0.005\\
    & $S_{1}$ &   0.0083 &  0.058 &  0.145 &  0.885 &  -0.105 &   0.121\\
    & $S_{2}$ &   0.2160 &  0.064 &  3.394 &  ** &   0.091 &   0.341\\
    & $S_{3}$ &   0.2577 &  0.068 &  3.807 &  *** &   0.125 &   0.390\\
    & $S_{4}$ &   0.1824 &  0.067 &  2.709 &  ** &   0.050 &   0.314\\
    & $S_{5}$ &   0.2959 &  0.067 &  4.414 &  *** &   0.164 &   0.427\\
    & $S_{6}$ &   0.3271 &  0.067 &  4.847 &  *** &   0.195 &   0.459\\
    & $S_{7}$ &   0.2751 &  0.064 &  4.307 &  *** &   0.150 &   0.400\\
    & $S_{8}$ &   0.2317 &  0.061 &  3.808 &  *** &   0.112 &   0.351\\
    & $S_{9}$ &   0.1735 &  0.056 &  3.075 &  ** &   0.063 &   0.284\\
    & $S_{10}$ &  0.1514 &  0.051 &  2.961 &  ** &   0.051 &   0.252\\
    & $S_{11}$ &  0.0506 &  0.038 &  1.335 &  0.182 &  -0.024 &   0.125\\
    \bottomrule
\end{tabular}
\\{\vspace{1mm}\hspace{2mm}\raggedright 
\footnotesize{Level of significance: * ($p < 0.05$), ** ($p < 0.01$), *** ($p < 0.001$).} \par}
\label{tab:nbr-composition-self-report}
\end{table}
\para{The Self-report Perspective} The global self-reported attack counts are collected weekly, so the dates do not exactly match the takedowns. The 1\textsuperscript{st} wave correlates with a statistically significant drop in total weekly attack counts from about 55M to 45M, followed by a gradual recovery over the subsequent few weeks (see \autoref{fig:self-reported-ddos-attacks} and \autoref{tab:nbr-composition-self-report}). The 2\textsuperscript{nd} wave's effects appeared minimal, which is consistent with other datasets. The Xmas'21 effect was not significant, similar to \amppot's view but contrary to \hopscotch's and \netscout's views.

Looking at self-reported statistics from individual booters, we found that a few weeks before the 1\textsuperscript{st} wave, several booters left the market voluntarily. The market structure did not follow the `monopoly effect' seen in the 2018 takedowns~\cite{collier2019booting}; the big booters did not absorb the market share of the seized ones, instead staying roughly the same size, growing in line with their previous trends. The two major booters (accounting for over 50\% of self-reported attacks) survived both takedowns; the recovery was due to smaller booters setting up, none of which captured significant market shares. After the 2\textsuperscript{nd} wave, one major booter captured the market share of some seized ones, but this did not lead to significant growth seen in 2018~\cite{collier2019booting}.

\para{The Overall Picture} While random artifacts cannot be entirely excluded, and the \hopscotch and \amppot datasets are inherently noisy (booters may not use all of the honeypots, and some attacks may not appear~\cite{griffioen2021scan}), comparing them and other datasets mitigates inference errors and provides a consistent picture: the 1\textsuperscript{st} wave had a statistically significant impact, while the 2\textsuperscript{nd} wave did not. However, the 1\textsuperscript{st} wave's impact, as observed by \hopscotch, \amppot, and \netscout, was significant on UDP-based attacks only but not on TCP-based attacks. As UDP protocols are commonly exploited by booters, this suggests an effect specifically on booters, while the overall long-term landscape was not significantly influenced. 

Xmas'21 appears to have mixed effects: the \hopscotch and \netscout views (UDP-based) are statistically significant, while the \amppot and self-report views are not. This differs from Xmas'22 (around the 1\textsuperscript{st} wave), when a significant impact is consistently seen across all datasets, especially as \netscout (the most stable one) shows UDP-based attacks reaching their lowest level, whereas the Xmas'21 drop (without takedowns) was milder. This suggests that the first-wave drop was likely associated with the takedown; however, declines in attacks around Christmas are also quite common~\cite{hiesgen2024age}. The impact may thus reflect a combined `upper bound' effect of all events, yet it remains short-lived; the takedown impact alone may be even less significant. Despite the 1\textsuperscript{st} wave's statistically significant impact, all DDoS datasets consistently indicate it was short-lived, lasting for at most about six weeks.

\para{Takeaways} The impact on the DDoS volume, especially UDP-based attacks, was statistically significant; however, the disruption was short-lived, lasting at most around six weeks.

\begin{figure}[t]
    \centering
    \includegraphics[width=0.475\textwidth]{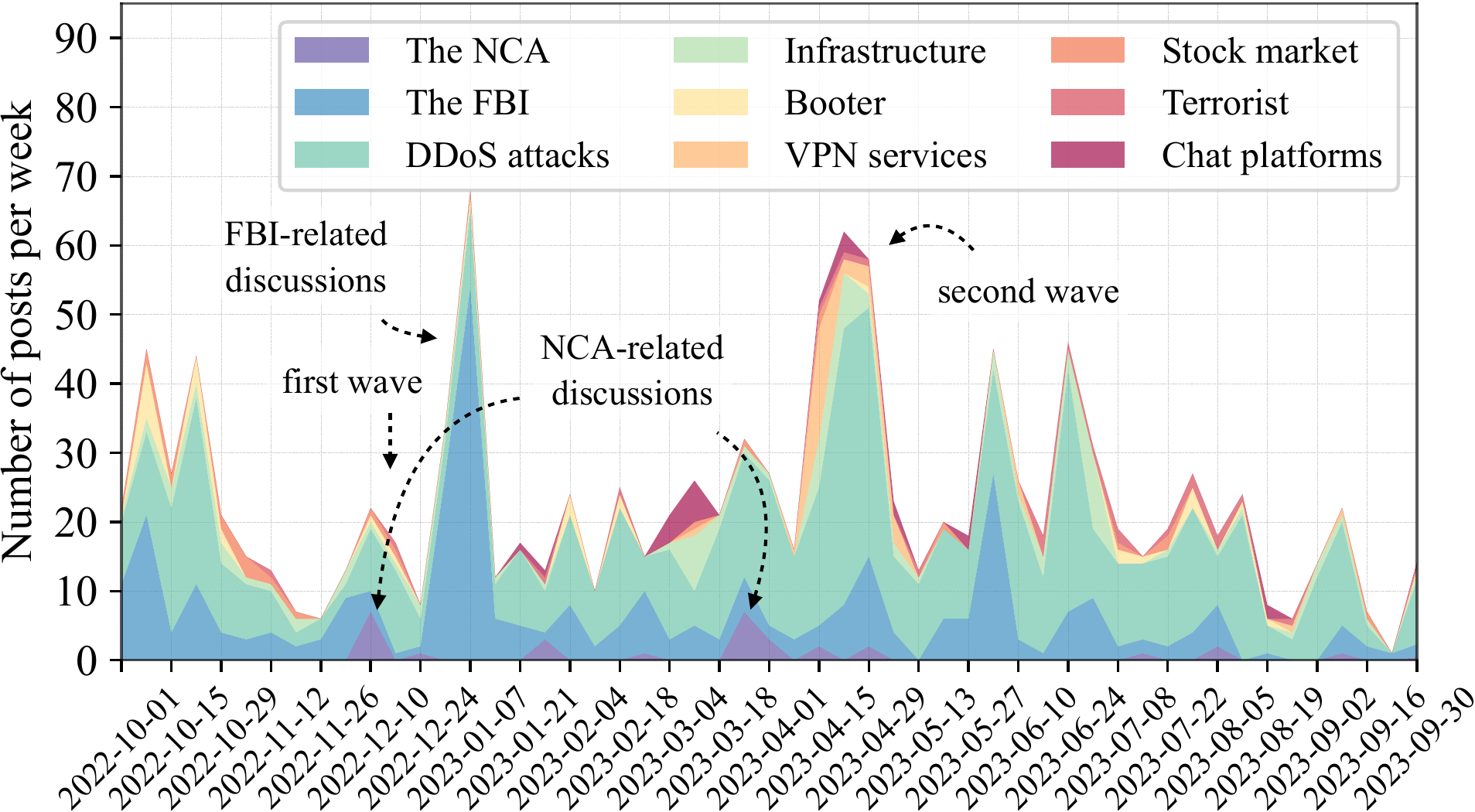}\\
    \caption{Number of weekly posts relevant to DDoS, booters, and the interventions on \hackforums by most popular topic.}
    \label{fig:hf-discussion-topics}
\end{figure}

\subsection{Community Perspectives}
To infer discussion topics, we train a BERTopic model~\cite{grootendorst2022bertopic} using \BTnAllRelevantPosts~\hackforums posts related to DDoS, booting, or interventions (with post content lemmatised before training). This model leverages transformer architecture for text embedding and provides a contextual semantics understanding, in some cases having better performance than Latent Dirichlet Allocation~\cite{blei2003latent}, especially for short content such as forum posts. We further qualitatively analyse key discussions related to takedowns and arrests to track the changing perceptions of users and booter operators. All quotes below were paraphrased to avoid direct searches that could reveal the posters.

\para{Discussion Topics} There was a sharp increase in FBI-related discussions around two weeks after the 1\textsuperscript{st} wave (see~\autoref{fig:hf-discussion-topics}). However, they largely pertained not to the takedowns but more broadly to the FBI in the media. There was also an increase in NCA-related posts, mainly directly in relation to the interventions. These were prompted by NCA-authored posts as part of their influence campaign. There are also some discussions and warnings about the NCA's deceptive tactics:

\begingroup
\addtolength\leftmargini{-0.15in}
\begin{quote}
    \quotetext{The NCA is apparently setting traps for cybercriminals by creating fake DDoS websites. It is wise to stay away from these sites completely, but these cybercriminals would still be making a lot of money, just like they have been always doing.}
\end{quote}
\endgroup

\noindent More recent NCA-related discussions on \hackforums focus on narratives of fear of arrest, with an increased perception of arrest likelihood for some crime types, including DDoS. Although small in number, they do reflect a genuine change in attitudes compared to assessments from five years prior~\cite{collier2019booting}, where narratives predominantly focused on the lack of skill and capacity of UK law enforcement. More broadly, there has been a clear effect on the perception of risk -- the narrative that these services are legal or ignored appears far less pervasive.

Most DDoS-related posts following the 1\textsuperscript{st} wave either reminded people not to mention DDoS or explicitly stated that DDoS is illegal, while a large proportion were related to protection services, VPNs, or new attack vectors. There were initially few posts about raids and takedowns, but this changed in early 2023 with rumours that a booter section might return to the forum with an alleged revision to the terms of service. 

\begingroup
\addtolength\leftmargini{-0.15in}
\begin{quote}
    \quotetext{Wow, that's news to me! I'd expect to see something like that in the marketplace, but I haven't come across it yet.} $\cdot$ 
    \quotetext{Yeah, same here. The documents currently have only a limited amount of prohibited stuff. I was absolutely certain that stressers would start pouring in.} $\cdot$ 
    \quotetext{Everyone still thinks they're banned.}
\end{quote}
\endgroup

\noindent There was an increase in discussions of booting following the 2\textsuperscript{nd} wave, which was sustained for a week but then dropped off significantly afterwards and continued to decline gradually. Interest in discussions about the FBI and the NCA also waned.

\begin{figure}[t]
    \centering
    \includegraphics[width=0.475\textwidth]{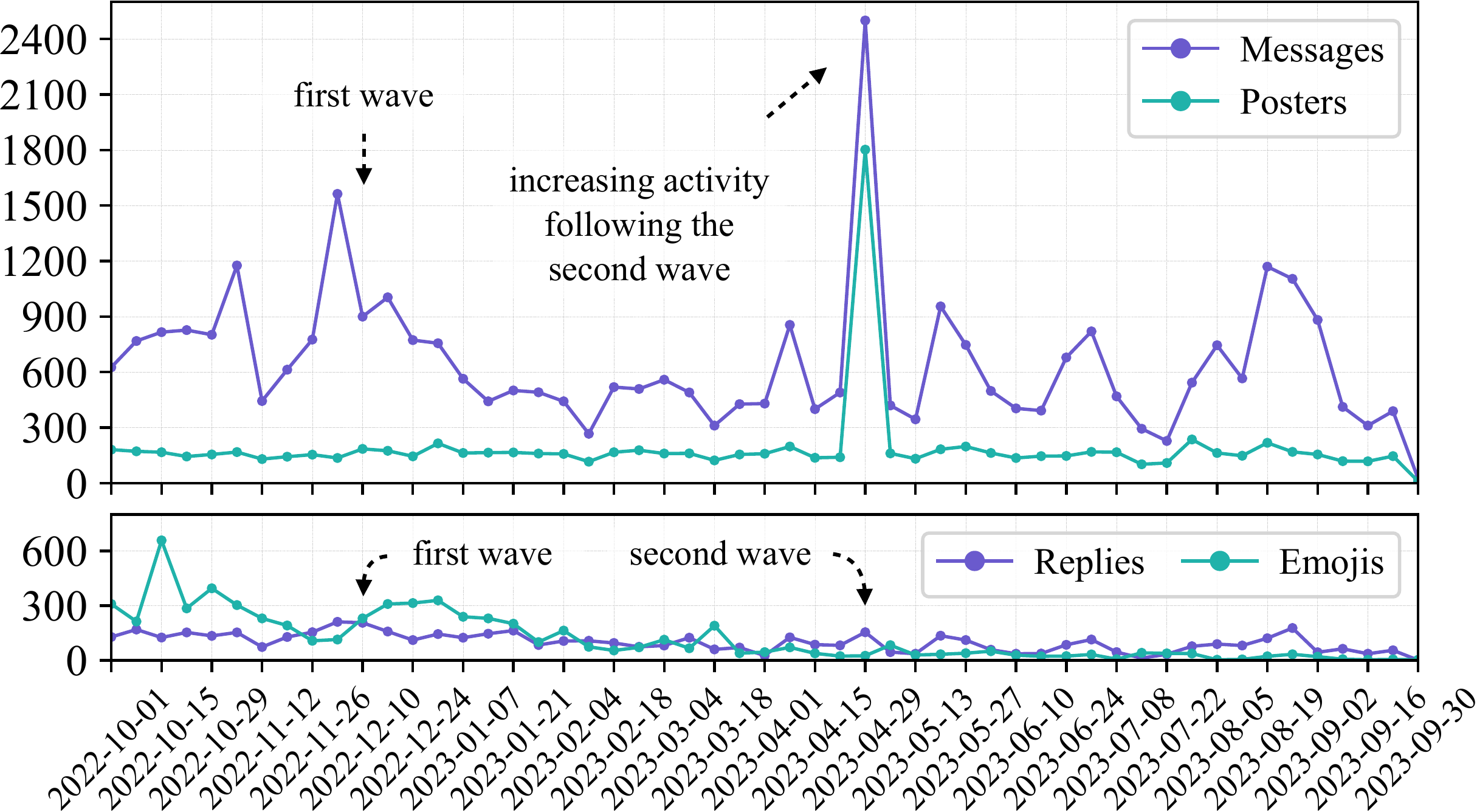}\\
    \caption{Number of messages, posters, and reactions of subscribers per week in all seized booters' Telegram channels.}
    \label{fig:telegram-activity}
\end{figure}

\para{Responses of Booter Users} The 1\textsuperscript{st} wave correlated with a two-month decline in the total number of messages on all seized booters' Telegram channels, dropping from around 1\,000 to 300 per week, while the number of posters (mostly the owners) was not affected (see~\autoref{fig:telegram-activity}). User reactions decreased from about 200 to 100 replies and from about 300 to less than 100 emojis after two months, indicating that users were less engaged. The 2\textsuperscript{nd} wave correlated with a sudden influx of posters and messages (mostly from new seized booters), but the impact on user reactions was mild, suggesting low engagement despite more users posting in the channels.

\para{Responses of Booter Operators} Some operators reacted quickly and attempted to resume operations, offering downtime compensation while their sites were being reconstructed.
\begingroup
\addtolength\leftmargini{-0.15in}
\begin{quote}
    \small
    \quotetext{We'll give you extra days on your plan due to unexpected downtime.} $\cdot$ 
    \quotetext{All clients received a one-day extension.}
\end{quote}
\endgroup

\noindent While all second-wave seized booters returned, 23 of the 48 first-wave seized booters did not. Among these, two explicitly gave up, advertising their domains and source code for sale or seeking freelance jobs; the rest quietly quit. Some of their Telegram chats indicate that they intended to leave the market.
\begingroup \addtolength\leftmargini{-0.15in}
\begin{quote} 
    \small
    \quotetext{Our current illegal income source is risky and unsustainable. police actions are happening, and many are coming in the future. [URL to the FBI seizure] -- 15 December 2022} $\cdot$
    \quotetext{Selling the source code of this booter, please message me -- 13 March 2023} $\cdot$
    \quotetext{Looking for a good developer? I'm currently available and ready to tackle your project, big or small. I have experience with Golang, Rust, C, C++, Python, JavaScript, and other languages. Don't hesitate to send me a message to discuss your needs. (Small projects are very cheap!) -- 19 May 2023}
\end{quote}
\endgroup

\para{Takeaways} There was seizure-related discussion in the hacking community; some changed their perception of risk. User engagement on booter channels dropped; some operators left.

\section{Discussion} \label{sec:discussion}
This concertedly conducted campaign is more extensive and persistent than the 2018 booter takedown~\cite{collier2019booting,kopp2019ddos} in some regards: (1) there have been two waves within a relatively short interval of four months, compared to the previous one wave; (2) more domains were seized (62 compared to 15 in 2018), along with more arrests; (3) deceptive sites were set up to attract users. Yet the effects still appear limited. It has triggered a classic cat-and-mouse game, with booter services repeatedly attempting to resurrect by announcing new domains through Telegram channels, while further takedowns may continue.

\para{The Efficacy} Prior analysis of the 2018 takedown~\cite{collier2019booting} mainly focuses on UDP amplification DDoS attacks, while our work covers multiple independent vantage points from both industry and academics, including ground-truth traffic. Our quantitative and qualitative analyses suggest that despite the relative fragility of the market in the beginning of the interventions, the overall effects appeared to be short-lived. The 1\textsuperscript{st} wave significantly disrupted the supply side, with half of the intervened booters eliminated. The substantial drops in traffic to seized booters itself is not surprising as users may leave quickly after noticing the seizures or registering accounts on the NCA's deceptive sites, but all resurrected domains attracted far less traffic than before, suggesting a positive takedown impact on the supply. UDP-based attacks (typically, though not exclusively attributed to booters) were largely impacted, with statistically significant drops observed across all four DDoS datasets. However, despite the intensive intervention efforts, it took only about six weeks for the attack volume to recover, about the same duration as in 2018~\cite{collier2019booting}. As seized booters could not fully resurrect, this rebound might be due to the emergence of brand new booters capturing market share, or users may use new services or attack vectors. The contrast between decreasing booter traffic and the recovering attack volume suggests that suppressing the supply alone may not suffice, as the demand likely persists in the long run. The 2\textsuperscript{nd} wave appeared to have minimal impact; all seized booters reappearing quickly with almost no major changes in attack volume. It may be that takedowns tend to cause market exit for only weakly-enrolled players, or a larger number of waves are required to disperse the committed remaining actors.

Though the effects of direct and indirect influence are generally impossible to directly attribute in an over-determined and complex social environment, there is evidence that perceptions of law enforcement have changed in the past five years. In particular, there is no longer significant evidence of a perception of booting as legal within the less technically skilled cybercrime community, for example on \hackforums, where interest and sentiment around these activities are generally declining. These communities, which are increasingly oriented around fraud and low-level scams, appear no longer to be a major customer base for booters. Instead, we see increased perceptions of risk, the waning of booters' popularity, and the increasing perceptions of law enforcement capacity. There is also evidence of booter operators exiting the market.

\para{Challenges} Unlike traditional crimes that are often confined to a single jurisdiction where law enforcement can be effective, the transnational nature of cybercrime slows down the process. Taking down bad sites is a long process involving complicated evidence gathering, legal procedures, and paperwork; its speed and effectiveness heavily depend on the incentives of requestors~\cite{moore2008impact}. However, the relief is often temporary; rapid recovery is possible as new domains and hosting can be set up within hours or days. Expired domains could be repurchased by bad actors to exploit the residual trust~\cite{moore2014ghosts,lever2016domain}. This also applies to seized domains: some are removed from the domain pool, but others are released and become repurchasable for malicious purposes~\cite{alowaisheq2019cracking}. Booters operating within the relevant jurisdictions could be seized, but if some infrastructure is moved to hidden services or other parts of the world, interventions might be less straightforward. 

Takedown, as a reactive strategy, does indeed have positive effects, but it might not be enough to fully address the issue in the long term, as also seen in the removal of phishing sites~\cite{moore2007examining}, the takedowns of botnets~\cite{nadji2013beheading,nadji2015still} and online cybercrime marketplaces~\cite{soska2015measuring}. Over a decade ago, millions of machines still remained infected years after Conficker, one of the largest botnets, was sinkholed, despite extensive cleanup efforts~\cite{asghari2015post}. Likewise, two years after the VPNFilter disruption, many routers were still compromised~\cite{vpnfiltertakedown}. More recently, the notorious LockBit ransomware resurfaced just one week after a coordinated intervention, Operation Cronos~\cite{lockbitresurface}. A drop in scans for default credentials of Ubiquiti routers coincided with the Moobot takedown, but it was not significant~\cite{moobottakedown}. 

Compared with larger-scale, more lucrative and culturally-embedded forms of crime (such as the global trade in narcotics), relatively small-scale law enforcement efforts in online markets can have the disruption and reshaping effects that require sustained action over decades and tens of billions of dollars annually to achieve in other markets~\cite{golz2018market}. Cybercriminals may sometimes get bored and `burn out'~\cite{collier2020cybercrime} and their interest can wane as has been seen for the volunteer hacktivists reacting to armed conflicts~\cite{vu2024getting,vu2023defacement}. It might be thought that industry could act more effectively than law enforcement~\cite{hutchings2016taking}, but when a series of swift and competent tech firms attempted to shut down an online hate and harassment forum in late 2022, it still recovered after a few months~\cite{vu2024no}. 

\section{Conclusion} \label{sec:conclusion}
Completely eliminating booters and DDoS attacks is hard; this issue persists and the market has proven characteristically resilient. Disruption efforts, such as takedowns, arrests, and legal threats, may function more as deterrents rather than long-lasting solutions. However, infrastructure takedowns, customer-facing influence interventions, and various ongoing industry actions since 2021 to tackle spoofing sources~\cite{collier2024peer}, especially when acting together, appear to have been effective in suppressing market growth in the short term. The combination of tactics we documented here seems to be becoming more widespread in dealing with persistent forms of cybercrime~\cite{collier2022influence} -- as also seen in the LockBit anti-ransomware actions recently taken by an international coalition of actors~\cite{lockbitintervention}. This reflects a more strategic and systematic approach by law enforcement for online crimes where jurisdiction renders arrest strategies untenable. Combining strategies, while unlikely to provoke a full market exit of all suppliers, is able to direct movement in practices, market structure, and organisation. By co-ordinating these and continuing to increase the friction on administrative practices, law enforcement could shape the moves made by these services, making them less accessible, less reliable, less advertisable, less trustworthy, and harder to run. While these recurring efforts will not stop committed actors from running and selling DDoS attacks in the long term, they contributes to making this as-a-service market untenable to operate at scale particularly during short periods of higher attack volumes such as school holidays and Christmas. Shifting such attacks from a volume crime market to a problem of skilled individuals would itself be a significant policy `win'.
\label{endofbody}

\section*{Acknowledgments} \label{sec:acknowledgments}
We are grateful to the U.S. Federal Bureau of Investigation, the U.K. National Crime Agency, and the Dutch Police for providing details about their interventions. We thank \netscout, the \ccc, and Christian Rossow for sharing their datasets. We are grateful to our colleagues Hugo Bijmans, Jack Hughes, Tina Marjanov, Yanna Papadodimitraki, Anna Talas, and Kieron Ivy Turk for their feedback. This work is supported by the European Research Council (ERC) under the European Union's Horizon 2020 research and innovation programme (grant agreement No 949127).

\appendix
\section{Ethics Considerations} \label{appendix:ethical-issues}
Our data collection and analysis were approved by our department's research ethics committee. We only scraped public forums and channels, which is lawful~\cite{webscrapinglegal}. We did not seek the consent of individuals on these forums and channels, as sending thousands of messages could be regarded as spamming. We expect users to be aware that their public postings are visible to all. Our analyses were conducted collectively to avoid individuals being identified, which accords with the British Society of Criminology’s Statement on Ethics~\cite{britishethics}. All quotes were paraphrased to prevent attribution. No domains of seized, functioning, or deceptive booters are disclosed to avoid attracting users, driving artificial traffic that confuses future collection, or causing potential harm to involved actors.

\section{Open Science} \label{appendix:data-licensing}
For the \netscout and \amppot datasets, please contact the providers. Datasets provided by the \ccc can be shared with researchers for appropriate projects through a license agreement~\cite{wilson2024identifying}. Datasets collected by us are also available, but their sensitive nature prevents them from being fully public. To simplify the process, our data is entrusted to the \ccc so that they can handle appropriate paperwork. They have a long history of data sharing in multiple jurisdictions and have established a proper regime in conjunction with legal specialists~\cite{cccagreement}.

\bibliographystyle{unsrt}
\bibliography{main}
\end{document}